\documentclass{article} %
\PassOptionsToPackage{usenames,dvipsnames}{xcolor}
\usepackage{iclr2022_conference,times}

\usepackage{amsmath,amsfonts,bm}

\def\eqref#1{equation~\ref{#1}}

\def\1{\bm{1}}

\DeclareMathAlphabet{\mathsfit}{\encodingdefault}{\sfdefault}{m}{sl}
\SetMathAlphabet{\mathsfit}{bold}{\encodingdefault}{\sfdefault}{bx}{n}

\usepackage{bbm}

\usepackage{hyperref}
\usepackage{url}

\usepackage{adjustbox}

\usepackage{algorithm}
\usepackage{algpseudocode}

\usepackage{wrapfig}

\usepackage{tikz}
\usepackage{pgfplots}
\DeclareUnicodeCharacter{2212}{−}
\usepgfplotslibrary{groupplots,dateplot}
\usetikzlibrary{patterns,shapes.arrows}
\pgfplotsset{compat=newest}
\usetikzlibrary{arrows}
\usepackage{url}
\usepackage{booktabs}

\usetikzlibrary{math}
\tikzmath
{
  function symlog(\x,\a){
    \yLarge = ((\x>\a) - (\x<-\a)) * (ln(max(abs(\x/\a),1)) + 1);
    \ySmall = (\x >= -\a) * (\x <= \a) * \x / \a ;
    return \yLarge + \ySmall ;
  };
  function symexp(\y,\a){
    \xLarge = ((\y>1) - (\y<-1)) * \a * exp(abs(\y) - 1) ;
    \xSmall = (\y>=-1) * (\y<=1) * \a * \y ;
    return \xLarge + \xSmall ;
  };
}

\usepackage[pdf]{graphviz}

\title{Does the Market of Citations Reward Reproducible Work?}

\author{Edward Raff \\
Booz Allen Hamilton\\
University of Maryland, Baltimore County \\
\texttt{raff\_edward@bah.com} 
}

\iclrfinalcopy %
\begin{document}

\maketitle

\begin{abstract}
The field of bibliometrics, studying citations and behavior, is critical to  the discussion of reproducibility. Citations are one of the primary incentive and reward systems for academic work, and so we desire to know if this incentive rewards reproducible work. Yet to the best of our knowledge, only one work has attempted to look at this combined space, concluding that non-reproducible work is more highly cited. 
We show that answering this question is more challenging than first proposed, and subtle issues can inhibit a robust conclusion. 
To make inferences with more robust behavior, we propose a hierarchical Bayesian model that incorporates the citation rate over time, rather than the total number of citations after a fixed amount of time. 
In doing so we show that, under current evidence  the answer is more likely that certain fields of study such as Medicine and Machine Learning (ML) do correlate reproducible works with more citations, but other fields appear to have no relationship. 
Further, we find that making code available and thoroughly referencing prior works appear to also positively correlate with increased citations. Our code and data can be found at \url{https://github.com/EdwardRaff/ReproducibleCitations}.
\end{abstract}

\section{Introduction}

A reproducibility crisis has been called for many scientific domains, including artificial intelligence and machine learning \citep{Donoho2009,Baker2016,Hutson725,Vul2008}. It is paramount that all disciplines work to remedy this situation and push for reproducible work both as good science, and to mitigate such crises. Such work has begun in various fields with different strategies~\citep{10.7554/eLife.67995,Poldrack2019,aac4716,10.5555/2969442.2969519,Gardner2018}, yet the incentive structure around producing reproducible work has received almost no attention. We note that the difference in terminology between reproduction and replicating is long, with conflicting terminology across fields and years \citep{Plesser2018}, we will use both terms interchangeably as our study focuses exclusively on cases where a different team independently performs the same experiments to obtain the same/similar results. 

Citations are the primary reward for academic outputs, and to our knowledge only the work of \cite{Serra-Garcia2021} has ever considered studying the relationship between papers that reproduce and the number of citations received. They used data on replication results from the fields of Psychology~\citep{aac4716}, Economics~\citep{RePEc:pra:mprapa:75461}, and Social Sciences~\citep{Camerer2018}. Distressingly, they conclude that non-reproducing work is cited more than reproducing works. 

Our work revisits this hypothesis and data, and draws a different conclusion. We will show in \autoref{sec:issues} that there are methodological issues that prevent a robust conclusion from being formed with the data and approach presented in \citep{Serra-Garcia2021}. Next, we will propose a Bayesian hierarchical model to alleviate these issues and allow further insight into the citation/replication question by incorporating a model of the citation rate changing over time in \autoref{sec:methodology}. In \autoref{sec:results} we show our model is a significantly better fit to the data, and concludes that citation rate is unrelated or positively correlated with reproduction success, depending on the field being studied. Finally, we will conclude in \autoref{sec:conclusion}.

\section{Related Work} \label{sec:related_work}

The study of paper citation has a long and multi-disciplinary history ~\citep{Lotka1926,Shockley1957,Price1965,https://doi.org/10.1002/asi.4630270505,Potter1981LotkasLR,Redner1998}, with many works proposing different power law variants to describe the distribution of citations. Most work that has looked at citations over time are looking at population level changes in citation distributions~\citep{Bornmann2015,Varga2019,Wallace2009}. We are aware of only one prior work that looked at the citation rate by year through studying the impact of publication-vs-arXiv~\citep{Traag2021}. This work also modeled citation rates as a Poisson, similar to \cite{Serra-Garcia2021}, which we will argue is an inappropriate model for citation count data. 

Used by \cite{Serra-Garcia2021} were \textit{negative citations}, a type of citation classification that can provide further insight into behaviors and results. The taxonomy of citation types, their labeling, and prediction \citep{Kunnath2022} are another lens through which insight may be gained, but is beyond the scope of our study. 

\citet{10.1145/1273496.1273526} produced one of the first applications of Bayesian modeling to the study of citation behavior and influences. Our task is different, and so our model bares little resemblance, but the overall strategy we argue is worth further study. Several latent factors exist in bibliometric study to which modern machine learning may yield benefits, and the scale of bibliometric data provides fertile ground to new and technical challenges to advance the field.

\section{Issues with Existing Modeling} \label{sec:issues}

While the Negative Binomial model has been previously identified to empirically have better performance at citation prediction \citep{Thelwall2014a}, the Poisson model is still very popular. We note though that there is an easier way to show the Poisson model is in fact, inappropriate, for the bibliometric research it is used. The Poisson model assumes the mean and variance are equal, and if the variance is larger than the mean, the model suffers from overdispersion that prevents meaningful results. A statistical test \citep{Cameron1990} confirms with $p< 0.001$ that this is the case for citation data, which in the data from \citep{Serra-Garcia2021} has a mean of 438 citations but a variance of 504,639.

While \citet{Serra-Garcia2021} used the Poisson model in their work on the connection between replication and reproducibility, we note there are additional factors that lead us to challenge their initial conclusion. The first is a data issue of reproducibility itself: $N=80$ documents were noted in \citep{Serra-Garcia2021}, but the data provide $N=139$ instances. We are unable to determine the correct selection criteria\footnote{The authors graciously spent considerable time working with us, and we did not have the same software licenses to use their saved results. One hypothesis from the authors was that non-significant results were excluded, but only removed 16 samples when we went through the data provided. Cross-discipline reproducibility and data sharing standards poses an interesting question beyond our scope.} to render only 80, and so proceed forward with the larger number of samples.

\begin{table}[!h]
\caption{Results indicating if successfully reproduced papers have more (positive) or less (negative) citations than papers that failed to reproduce. Models tested include Poisson verse NegativeBinomial (NB) regressions using the original three domains with Google Scholar (GS) or Semantic Scholar (SC) citations each, and an additional case using SC with a fourth set of reproduction results from the Medical domain (+M). }\label{tbl:rpd}
    \adjustbox{max width=\columnwidth}{%
\begin{tabular}{@{}cclclclclclcl@{}}
\toprule
                               & \multicolumn{2}{c}{Poisson-GS}                     & \multicolumn{2}{c}{Poisson-SC}                     & \multicolumn{2}{c}{Poisson-SC+M}                   & \multicolumn{2}{c}{NB-GS}                          & \multicolumn{2}{c}{NB-SC}                          & \multicolumn{2}{c}{NB-SC+M}                        \\ \cmidrule(lr){2-3}\cmidrule(lr){4-5} \cmidrule(lr){6-7} \cmidrule(lr){8-9} \cmidrule(lr){10-11} \cmidrule(lr){12-13}
                               & coef                       & \multicolumn{1}{c}{p} & coef                       & \multicolumn{1}{c}{p} & coef                       & \multicolumn{1}{c}{p} & coef                       & \multicolumn{1}{c}{p} & coef                       & \multicolumn{1}{c}{p} & coef                       & \multicolumn{1}{c}{p} \\ \midrule
\multicolumn{1}{l}{Reproduced} & \multicolumn{1}{l}{0.0172} & 0.129                 & \multicolumn{1}{l}{0.1138} & $<$0.001                & \multicolumn{1}{l}{0.5775} & $<$0.001                & \multicolumn{1}{l}{0.0172} & 0.150                 & \multicolumn{1}{l}{4.4592} & $<$0.001                & \multicolumn{1}{l}{0.5777} & 0.004                 \\ \bottomrule
\end{tabular}
}
\end{table}

To demonstrate the lack of robustness to the prior methodology, we will perform several repetitions of the overall approach choosing between:

\begin{enumerate}
    \item Using the Poisson model versus a Negative-Binomial model
    \item Using the original Google Scholar (GS) citation count data provided vs citation data from Semantic Scholar (SC)
    \item Using the original data with (SC) additionally with reproduction results from the Medical domain, adding a fourth field (+M). 
\end{enumerate}

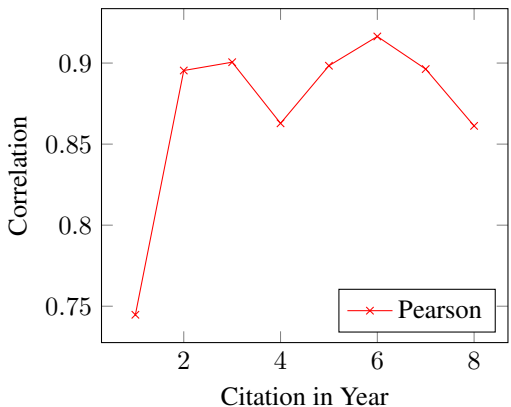
\begin{wrapfigure}{R}{0.51\textwidth}
    \centering
    \begin{tikzpicture}
	\begin{axis}[
	    legend pos=south east,
	    width=0.5\textwidth,
		xlabel=Citation in Year,
		ylabel=Correlation]
	\addplot[color=red,mark=x] coordinates {
	(1, 0.7447260788124388)
	(2, 0.8954058003831814)
	(3, 0.9005417432391989)
	(4, 0.8628314100903861)
	(5, 0.8983553314300697)
	(6, 0.9164083066431671)
	(7, 0.8962944112597322)
	(8, 0.8612543345660336)
	};
	\addlegendentry{Pearson}
	\end{axis}
\end{tikzpicture}
    \caption{Correlation between Google Scholar and Semantic Scholar in the number of citations for each document per year. After multiple-test correction all years were significantly correlated with $p<0.001$ in all cases. }
    \label{fig:correlation}
\end{wrapfigure}

This provides six total results, presented in \autoref{tbl:rpd} using \citep{seabold2010statsmodels}. We can see in that no case do we observe a negative indication that papers which fail to replicate are cited more. However, we do see inconsistent conclusion about the impact of replication itself. When using Google Scholar the conclusion is there is no relationship, and when using Semantic Scholar the conclusion is a strong relationship. This challenge is not a factor of these citations sources having dramatic disagreement, as can be seen in \autoref{fig:correlation} both are highly correlated in the per-year citations of the documents. This issue is instead that of model fit, as the highest adjusted $R^2$ fit amongst the Negative Binomial models is $0.0039$. 

\begin{figure}[!h]
    \centering
    \adjustbox{max width=0.48\columnwidth}{%
    \begin{tikzpicture}

\definecolor{brown1462651}{RGB}{146,26,51}
\definecolor{darkgray178}{RGB}{178,178,178}
\definecolor{darkolivegreen7110943}{RGB}{71,109,43}
\definecolor{darkslategray46}{RGB}{46,46,46}
\definecolor{lightgray204}{RGB}{204,204,204}
\definecolor{silver188}{RGB}{188,188,188}
\definecolor{slategray125111158}{RGB}{125,111,158}
\definecolor{steelblue69133171}{RGB}{69,133,171}
\definecolor{whitesmoke238}{RGB}{238,238,238}

\begin{axis}[
axis background/.style={fill=whitesmoke238},
axis line style={silver188},
legend cell align={left},
legend style={
  fill opacity=0.8,
  draw opacity=1,
  text opacity=1,
  at={(1,0.5)},
  anchor=west,
  draw=lightgray204,
  fill=whitesmoke238
},
log basis y={10},
tick pos=left,
title={Citations By Field},
x grid style={darkgray178},
width=0.48\textwidth,
height=0.48\textwidth,
xlabel={Replicated},
xmin=-0.5, xmax=1.5,
xtick style={color=black},
xtick={0,1},
xticklabels={No,Yes},
y grid style={darkgray178},
ylabel={Total Citations},
ymajorgrids,
ymin=1, ymax=3771.95533198895,
ymode=log,
ytick style={color=black},
ytick={0.1,1,10,100,1000,10000,100000},
yticklabels={
  \(\displaystyle {10^{-1}}\),
  \(\displaystyle {10^{0}}\),
  \(\displaystyle {10^{1}}\),
  \(\displaystyle {10^{2}}\),
  \(\displaystyle {10^{3}}\),
  \(\displaystyle {10^{4}}\),
  \(\displaystyle {10^{5}}\)
}
]
\path [draw=darkslategray46, fill=steelblue69133171, thick]
(axis cs:-0.398,68)
--(axis cs:-0.202,68)
--(axis cs:-0.202,156.5)
--(axis cs:-0.398,156.5)
--(axis cs:-0.398,68)
--cycle;
\path [draw=darkslategray46, fill=brown1462651, thick]
(axis cs:-0.198,19)
--(axis cs:-0.002,19)
--(axis cs:-0.002,69)
--(axis cs:-0.198,69)
--(axis cs:-0.198,19)
--cycle;
\path [draw=darkslategray46, fill=slategray125111158, thick]
(axis cs:0.00200000000000003,131.5)
--(axis cs:0.198,131.5)
--(axis cs:0.198,380.75)
--(axis cs:0.00200000000000003,380.75)
--(axis cs:0.00200000000000003,131.5)
--cycle;
\path [draw=darkslategray46, fill=darkolivegreen7110943, thick]
(axis cs:0.202,504)
--(axis cs:0.398,504)
--(axis cs:0.398,731)
--(axis cs:0.202,731)
--(axis cs:0.202,504)
--cycle;
\path [draw=darkslategray46, fill=steelblue69133171, thick]
(axis cs:0.602,46.5)
--(axis cs:0.798,46.5)
--(axis cs:0.798,146)
--(axis cs:0.602,146)
--(axis cs:0.602,46.5)
--cycle;
\path [draw=darkslategray46, fill=brown1462651, thick]
(axis cs:0.802,20.5)
--(axis cs:0.998,20.5)
--(axis cs:0.998,74.5)
--(axis cs:0.802,74.5)
--(axis cs:0.802,20.5)
--cycle;
\path [draw=darkslategray46, fill=slategray125111158, thick]
(axis cs:1.002,130.5)
--(axis cs:1.198,130.5)
--(axis cs:1.198,209.5)
--(axis cs:1.002,209.5)
--(axis cs:1.002,130.5)
--cycle;
\path [draw=darkslategray46, fill=darkolivegreen7110943, thick]
(axis cs:1.202,632.25)
--(axis cs:1.398,632.25)
--(axis cs:1.398,1364.25)
--(axis cs:1.202,1364.25)
--(axis cs:1.202,632.25)
--cycle;
\draw[draw=darkslategray46,fill=steelblue69133171] (axis cs:0,0) rectangle (axis cs:0,0);
\addlegendimage{ybar,ybar legend,draw=darkslategray46,fill=steelblue69133171}
\addlegendentry{Economics}

\draw[draw=darkslategray46,fill=brown1462651] (axis cs:0,0) rectangle (axis cs:0,0);
\addlegendimage{ybar,ybar legend,draw=darkslategray46,fill=brown1462651}
\addlegendentry{Psychology}

\draw[draw=darkslategray46,fill=slategray125111158] (axis cs:0,0) rectangle (axis cs:0,0);
\addlegendimage{ybar,ybar legend,draw=darkslategray46,fill=slategray125111158}
\addlegendentry{Social}

\draw[draw=darkslategray46,fill=darkolivegreen7110943] (axis cs:0,0) rectangle (axis cs:0,0);
\addlegendimage{ybar,ybar legend,draw=darkslategray46,fill=darkolivegreen7110943}
\addlegendentry{Medicine}

\addplot [thick, darkslategray46, forget plot]
table {%
-0.3 68
-0.3 20
};
\addplot [thick, darkslategray46, forget plot]
table {%
-0.3 156.5
-0.3 166
};
\addplot [thick, darkslategray46, forget plot]
table {%
-0.349 20
-0.251 20
};
\addplot [thick, darkslategray46, forget plot]
table {%
-0.349 166
-0.251 166
};
\addplot [black, mark=diamond*, mark size=2.5, mark options={solid,fill=darkslategray46}, only marks, forget plot]
table {%
-0.3 358
};
\addplot [thick, darkslategray46, forget plot]
table {%
-0.1 19
-0.1 1
};
\addplot [thick, darkslategray46, forget plot]
table {%
-0.1 69
-0.1 138
};
\addplot [thick, darkslategray46, forget plot]
table {%
-0.149 1
-0.051 1
};
\addplot [thick, darkslategray46, forget plot]
table {%
-0.149 138
-0.051 138
};
\addplot [black, mark=diamond*, mark size=2.5, mark options={solid,fill=darkslategray46}, only marks, forget plot]
table {%
-0.1 213
-0.1 150
};
\addplot [thick, darkslategray46, forget plot]
table {%
0.1 131.5
0.1 115
};
\addplot [thick, darkslategray46, forget plot]
table {%
0.1 380.75
0.1 591
};
\addplot [thick, darkslategray46, forget plot]
table {%
0.051 115
0.149 115
};
\addplot [thick, darkslategray46, forget plot]
table {%
0.051 591
0.149 591
};
\addplot [thick, darkslategray46, forget plot]
table {%
0.3 504
0.3 347
};
\addplot [thick, darkslategray46, forget plot]
table {%
0.3 731
0.3 838
};
\addplot [thick, darkslategray46, forget plot]
table {%
0.251 347
0.349 347
};
\addplot [thick, darkslategray46, forget plot]
table {%
0.251 838
0.349 838
};
\addplot [black, mark=diamond*, mark size=2.5, mark options={solid,fill=darkslategray46}, only marks, forget plot]
table {%
0.3 19
0.3 1187
0.3 1195
};
\addplot [thick, darkslategray46, forget plot]
table {%
0.7 46.5
0.7 1
};
\addplot [thick, darkslategray46, forget plot]
table {%
0.7 146
0.7 193
};
\addplot [thick, darkslategray46, forget plot]
table {%
0.651 1
0.749 1
};
\addplot [thick, darkslategray46, forget plot]
table {%
0.651 193
0.749 193
};
\addplot [thick, darkslategray46, forget plot]
table {%
0.9 20.5
0.9 6
};
\addplot [thick, darkslategray46, forget plot]
table {%
0.9 74.5
0.9 138
};
\addplot [thick, darkslategray46, forget plot]
table {%
0.851 6
0.949 6
};
\addplot [thick, darkslategray46, forget plot]
table {%
0.851 138
0.949 138
};
\addplot [black, mark=diamond*, mark size=2.5, mark options={solid,fill=darkslategray46}, only marks, forget plot]
table {%
0.9 171
0.9 162
};
\addplot [thick, darkslategray46, forget plot]
table {%
1.1 130.5
1.1 51
};
\addplot [thick, darkslategray46, forget plot]
table {%
1.1 209.5
1.1 259
};
\addplot [thick, darkslategray46, forget plot]
table {%
1.051 51
1.149 51
};
\addplot [thick, darkslategray46, forget plot]
table {%
1.051 259
1.149 259
};
\addplot [black, mark=diamond*, mark size=2.5, mark options={solid,fill=darkslategray46}, only marks, forget plot]
table {%
1.1 373
};
\addplot [thick, darkslategray46, forget plot]
table {%
1.3 632.25
1.3 1
};
\addplot [thick, darkslategray46, forget plot]
table {%
1.3 1364.25
1.3 1732
};
\addplot [thick, darkslategray46, forget plot]
table {%
1.251 1
1.349 1
};
\addplot [thick, darkslategray46, forget plot]
table {%
1.251 1732
1.349 1732
};
\addplot [thick, darkslategray46, forget plot]
table {%
-0.398 142
-0.202 142
};
\addplot [thick, darkslategray46, forget plot]
table {%
-0.198 35
-0.002 35
};
\addplot [thick, darkslategray46, forget plot]
table {%
0.00200000000000003 235.5
0.198 235.5
};
\addplot [thick, darkslategray46, forget plot]
table {%
0.202 610
0.398 610
};
\addplot [thick, darkslategray46, forget plot]
table {%
0.602 76
0.798 76
};
\addplot [thick, darkslategray46, forget plot]
table {%
0.802 39
0.998 39
};
\addplot [thick, darkslategray46, forget plot]
table {%
1.002 161
1.198 161
};
\addplot [thick, darkslategray46, forget plot]
table {%
1.202 982
1.398 982
};
\end{axis}

\end{tikzpicture}
    }
    \adjustbox{max width=0.48\columnwidth}{%
    \input{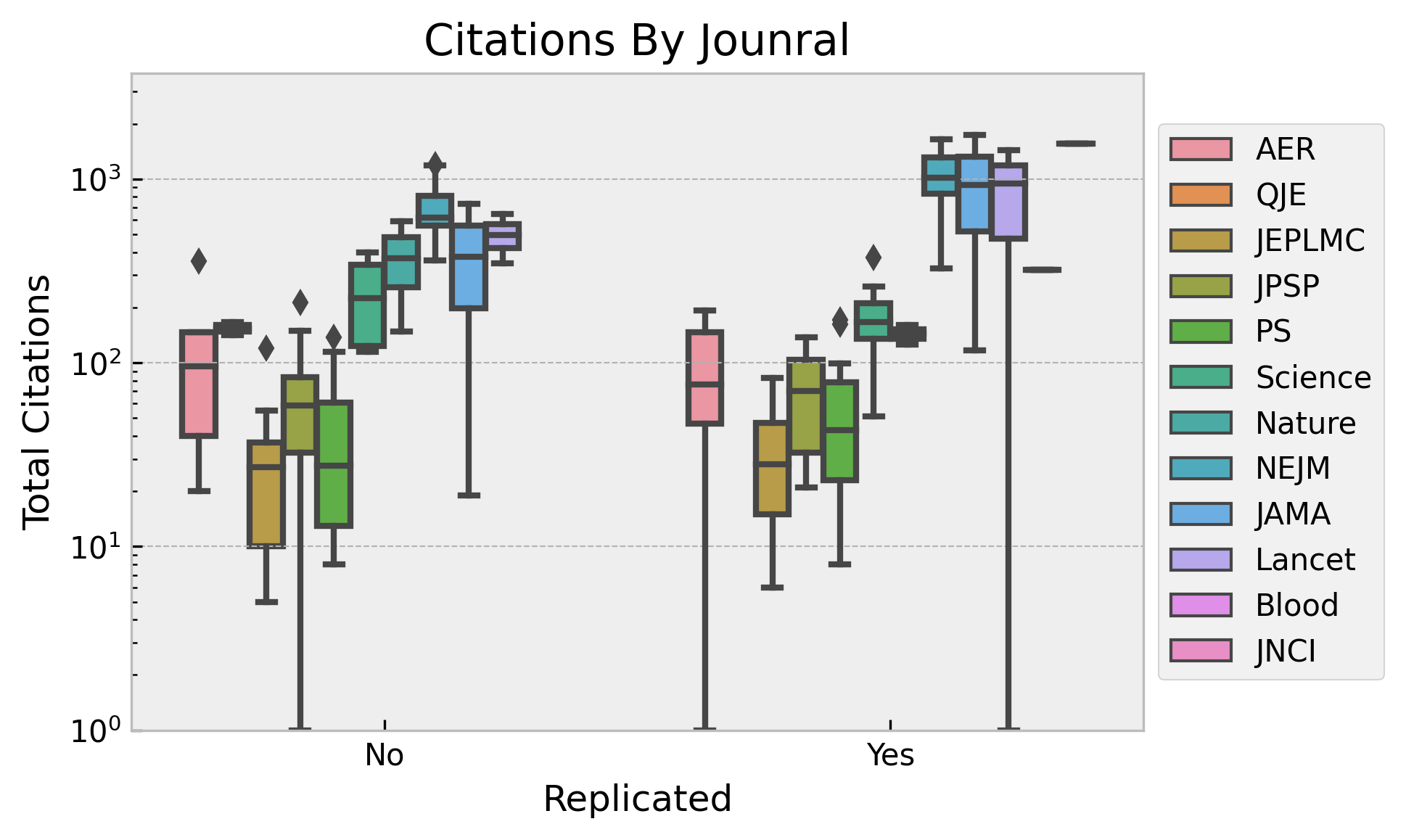}
    }
    \caption{Total number of citations accumulated for replicated and failed to replicate papers grouped by field (left) and journal (right).}
    \label{fig:citations_by_source}
\end{figure}
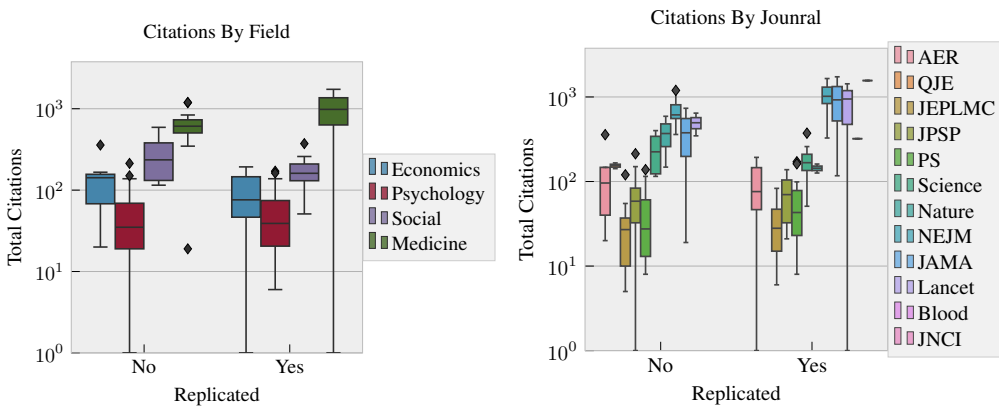

The source of this discrepancy is inappropriate merging of all data sources into one pool. The papers selected from Economics, Psychology, Social Science, and Medicine where all selected with biases toward higher citation rates --- largely through selection of high impact factor sources. The citation rate per field, or journal, are not the same, as shown in \autoref{fig:citations_by_source}. Imbalances in the number of papers from each source that happened to replicate or not amplify spurious noise, resulting in low model fit and unstable conclusions.

\section{Methodology} \label{sec:methodology}

To address these problems, we propose a Bayesian hierarchical model that incorporates the citation rate over time, rather than the cumulative total number of citations. Our interest in citation rate over time is of interest not merely for model fit, but primarily because we are interested if the types of citation patterns vary between reproducible and non-reproducible papers. That is to say, some papers do not start to accumulate citations for a considerable amount of time, others reach a steady-state of citations, and others reach a peak citation rate before their citation rate drops. A total-citation rate model can not reveal anything about this question. 

\begin{figure}[!h]
    \centering
    \adjustbox{max width=\columnwidth}{%
    \includegraphics{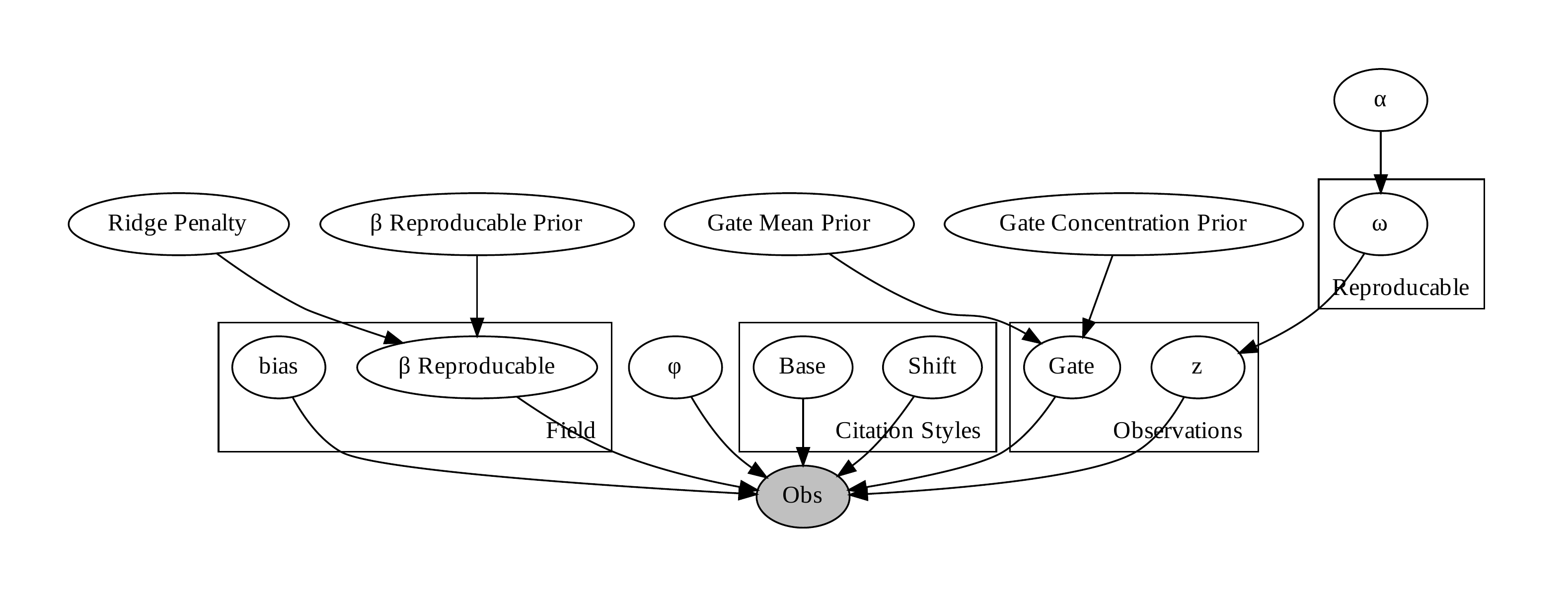}
    }
    \caption{Plate diagram of our proposed citation-replicated model. The observations are done against a Negative-Binomial model. }
    \label{fig:plate_sci}
\end{figure}

The high level plate diagram of our approach is presented in \autoref{fig:plate_sci}, which we will discuss at a high level with the detailed generative story given by \autoref{algo:model}. The coefficients $\beta$ are with respect to each Field, with a hierarchical prior used over then and a shared ridge regression penalty (variance of the Gaussian distribution). 

\begin{equation} \label{eq:neg_binomal2}
    \text{NegBinomial2}(n \, | \, \mu, \varphi)  = \binom{n + \varphi - 1}{n} \,
\left( \frac{\mu}{\mu+\varphi} \right)^{\!n} \, \left(
\frac{\varphi}{\mu+\varphi} \right)^{\!\varphi} \!.
\end{equation}

The observations are done with respect to a zero-inflated Negative-Binomial model, parameterized with a mean and dispersion factor $\mu$ and $\varphi$ as shown in \autoref{eq:neg_binomal2}. The zero-inflation serves two purposes. First, some papers do receive zero citations for some time before becoming popular, and the zero-inflation model prevents down-weighting the citation rate $\mu$ from these zero citations. Second, it allows us a convenient way to handle the fact that papers were published at different times, and thus for a desired horizon of $T$ years not all papers will have $T$ years of existence to accumulate citations. When a year has not yet occurred, we force the zero inflation gate to effectively mask the year with no impact on the model. We used a target of $T=10$ years in all cases. Each paper receives it's own gate value with a hyper prior shared over all samples. We use the proportional Beta hyper prior as shown in \autoref{eq:beta_prop} with a non-informative prior over $\mu$. 

\begin{equation} \label{eq:beta_prop}
\text { BetaProportion }(\theta \mid \mu, \kappa)=\frac{1}{\mathrm{~B}(\mu \kappa,(1-\mu) \kappa)} \theta^{\mu \kappa-1}(1-\theta)^{(1-\mu) \kappa-1}
\end{equation}

To represent the impact of the $t$'th year's citation rate of the $i$'th sample $\mu_{i,t}$ we model a base citation rate $\mu_i$ modulated by an annual \textit{base} citation multiplier sampled from a Gamma prior centered at a mean of 1.0 (i.e., no change in annual citation rate). The impact of the compounding base rate can be delayed (but not increased, as that implies pre-publication citations) by a \textit{shift} factor samples from a positive Laplacian scaled so that the entire $T$ years may be selected by the prior would prefer no shift. 

We do not give each sample it's own base and shift as it allows significant over-fitting of the model to ignore the impact of the coefficients $\beta$. Instead we use a Dirichlet process to sample from a pool base/shift pairs --- where reproducible and non-reproducible papers each receive a separate Dirichlet process sampling from the same pool. We enforce a sparse process by putting a Beta prior over the $\alpha$ parameter of the processes so that we may see if there is a difference in the types of citation styles between papers (e.g., do non-replicating papers more frequently have decaying base rates $<$ 1). In each experiment there is one pool of base/shift pairs, and two sets of distributions $\omega$ over those pools. One  $\omega_S$ for reproduced papers and one $\omega_F$ for the non-reproduced. In this way the model can inform us if there appears to be a difference ($\omega_S \neq \omega_F$) in citation styles (base/shift pairs) between the populations.

\begin{algorithm}[!h]
\caption{Our Hierarchical Bayesian generative story for modeling citation rates. The ${}^+$ indicates distributions truncated to be non-negative.  } \label{algo:model}
\begin{algorithmic}
\Require $N$ observations with $r_i \in \{S, F\}$ for successful or failed reproduction and $f_i$ indicating the field of research for the paper. 
\State $\lambda_\mathit{ridge} \sim $ HalfCauchy(0,1)
\State $\alpha \sim $ Beta(1, 10) \Comment{A Beta distribution used to encourage sparse solutions}
\State $\omega^S \sim $ Dirichlet($\alpha$) \Comment{A different distribution over all base/shift values for reproducible $\ldots$ }
\State $\omega^F \sim $ Dirichlet($\alpha$) \Comment{and non-reproducible papers}
\ForAll{$i \in {1, \ldots, \infty}$} \Comment{Citation Styles for $\omega^*$ will sample from }
    \State $\mathit{shift} \sim \text{Laplace}^+(0, \text{years out}/6)$
    \State $\mathit{base} \sim \Gamma(100, 100)$ \Comment{This Gamma distribution will encourage values near 1, as values > 2 are undesirable in being unrealistic. }
\EndFor
\State $\widehat{\beta^\mathit{field}} \sim \mathcal{N}(0, 1)$ \Comment{Hierarchical Reproducible Prior}
\ForAll{Field of Study $i$}
    \State $\beta^\mathit{field}_i \sim \mathcal{N}(\widehat{\beta^\mathit{field}}, \lambda_\mathit{ridge})$
    \State $b_i \sim $ Cauchy(0, 1) \Comment{Bias term is independent between Fields}
\EndFor
\State $\widehat{\text{gate}^\mu} \sim \mathcal{U}(0, 1)$ \Comment{Uninformative prior on the mean rate of no citations occurring.}
\State $\widehat{\text{gate}^\kappa} \sim \Gamma(1, 20)$
\State $\varphi \sim \text{Cauchy}^+(0, 5)$
\ForAll{Observations $i$}
    \State $z \sim $ Categorical($\omega^{r_i}$) \Comment{Select the citation style base/shift for this sample based on the distribution w.r.t. the sample replicating or not}
    \State $\log(\mu_i) \gets \beta^\mathit{field}_{f_i} \cdot \mathbbm{1}[r_i = S] + b_{f_i} $ \Comment{The rate is modified based on the paper replicating or not.}
    \State $\mathit{gate}_i \sim \text{BetaProportion}(\widehat{\text{gate}^\mu}, \widehat{\text{gate}^\kappa})$
    \ForAll{Time steps $t$}
        \State $\mu_{i,t} \gets \mu_i \cdot \mathit{base}_z^{\max(t-\mathit{shift}_z, 0)}$
        \State accumulate Zero-Inflated Negative Binomial loss NetBinomial2($y_i | \mu_{i,t}, \varphi$) with gate probability $\mathit{gate}_i$
    \EndFor
\EndFor
\end{algorithmic}
\end{algorithm}

 The full model is detailed in \autoref{algo:model}. We use NumPyro~\citep{Phan2019} to implement the model with the NUTS sampler~\citep{JMLR:v15:hoffman14a}. In all cases we use 500 burn-in iterations followed by 2,250 steps with a thinning factor of 3. 

\section{Results} \label{sec:results}

Now that we have specified our approach to understanding how citations may be impacted by a paper's ability to replicate, we will present out results in two sections. First we will consider the results with respect to the previous fields of study (Medicine, Economics, Psychology, Social) and show that we obtain consistent results and reasonably believe them to be a more reliable model. Second we will repeat the study applied to data from machine learning~\citep{Raff2019_quantify_repro}. This data is studied separately because it has a different kind of selection bias, and a different set of available features to consider, than the other data. 

\subsection{Science Results}

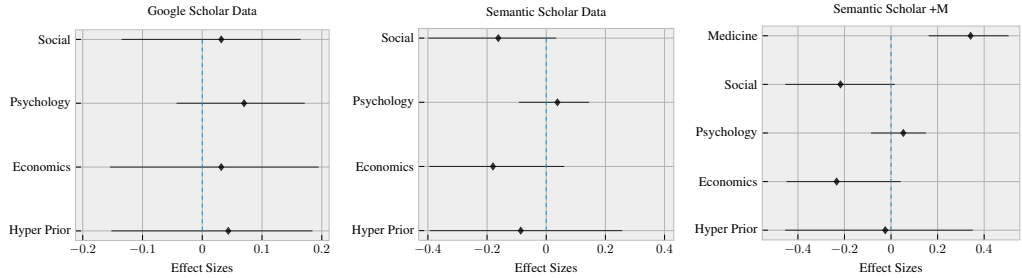
\begin{figure}[!h]
    \centering
    \adjustbox{max width=0.32\columnwidth}{%
    \begin{tikzpicture}

\definecolor{darkgray178}{RGB}{178,178,178}
\definecolor{silver188}{RGB}{188,188,188}
\definecolor{steelblue52138189}{RGB}{52,138,189}
\definecolor{whitesmoke238}{RGB}{238,238,238}

\begin{axis}[
axis background/.style={fill=whitesmoke238},
axis line style={silver188},
tick pos=left,
title={Google Scholar Data},
x grid style={darkgray178},
xlabel={Effect Sizes},
xmajorgrids,
xmin=-0.212291495501995, xmax=0.212291495501995,
xtick style={color=black},
y grid style={darkgray178},
ymajorgrids,
ymin=-0.15, ymax=3.15,
ytick style={color=black},
ytick={0,1,2,3},
yticklabels={
  Hyper Prior,
  Economics,
  Psychology,
  Social
}
]
\path [draw=black, draw opacity=0.8, thick]
(axis cs:-0.152138784527779,0)
--(axis cs:0.184362322092056,0);

\path [draw=black, draw opacity=0.8, thick]
(axis cs:-0.154349520802498,1)
--(axis cs:0.194832399487495,1);

\path [draw=black, draw opacity=0.8, thick]
(axis cs:-0.0430942252278328,2)
--(axis cs:0.171511396765709,2);

\path [draw=black, draw opacity=0.8, thick]
(axis cs:-0.134997844696045,3)
--(axis cs:0.164590314030647,3);

\addplot [thick, steelblue52138189, dashed]
table {%
0 0
0 1
0 2
0 3
};
\addplot [thick, black, opacity=0.8, mark=diamond*, mark size=2.5, mark options={solid}, only marks]
table {%
0.0435357876121998 0
0.0317949093878269 1
0.0700091794133186 2
0.031797107309103 3
};
\end{axis}

\end{tikzpicture}
    }
    \adjustbox{max width=0.32\columnwidth}{%
    \begin{tikzpicture}

\definecolor{darkgray178}{RGB}{178,178,178}
\definecolor{silver188}{RGB}{188,188,188}
\definecolor{steelblue52138189}{RGB}{52,138,189}
\definecolor{whitesmoke238}{RGB}{238,238,238}

\begin{axis}[
axis background/.style={fill=whitesmoke238},
axis line style={silver188},
tick pos=left,
title={Semantic Scholar Data},
x grid style={darkgray178},
xlabel={Effect Sizes},
xmajorgrids,
xmin=-0.431188015639782, xmax=0.431188015639782,
xtick style={color=black},
y grid style={darkgray178},
ymajorgrids,
ymin=-0.15, ymax=3.15,
ytick style={color=black},
ytick={0,1,2,3},
yticklabels={
  Hyper Prior,
  Economics,
  Psychology,
  Social
}
]
\path [draw=black, draw opacity=0.8, thick]
(axis cs:-0.394055128097534,0)
--(axis cs:0.257005274295807,0);

\path [draw=black, draw opacity=0.8, thick]
(axis cs:-0.395296692848206,1)
--(axis cs:0.0609618425369263,1);

\path [draw=black, draw opacity=0.8, thick]
(axis cs:-0.0924994647502899,2)
--(axis cs:0.145027905702591,2);

\path [draw=black, draw opacity=0.8, thick]
(axis cs:-0.39841690659523,3)
--(axis cs:0.0337300449609756,3);

\addplot [thick, steelblue52138189, dashed]
table {%
0 0
0 1
0 2
0 3
};
\addplot [thick, black, opacity=0.8, mark=diamond*, mark size=2.5, mark options={solid}, only marks]
table {%
-0.086047425866127 0
-0.179994568228722 1
0.0377715826034546 2
-0.16207417845726 3
};
\end{axis}

\end{tikzpicture}
    }
    \adjustbox{max width=0.32\columnwidth}{%
    \begin{tikzpicture}

\definecolor{darkgray178}{RGB}{178,178,178}
\definecolor{silver188}{RGB}{188,188,188}
\definecolor{steelblue52138189}{RGB}{52,138,189}
\definecolor{whitesmoke238}{RGB}{238,238,238}

\begin{axis}[
axis background/.style={fill=whitesmoke238},
axis line style={silver188},
tick pos=left,
title={Semantic Scholar +M},
x grid style={darkgray178},
xlabel={Effect Sizes},
xmajorgrids,
xmin=-0.55331034809351, xmax=0.55331034809351,
xtick style={color=black},
y grid style={darkgray178},
ymajorgrids,
ymin=-0.2, ymax=4.2,
ytick style={color=black},
ytick={0,1,2,3,4},
yticklabels={
  Hyper Prior,
  Economics,
  Psychology,
  Social,
  Medicine
}
]
\path [draw=black, draw opacity=0.8, thick]
(axis cs:-0.454829663038254,0)
--(axis cs:0.351390808820724,0);

\path [draw=black, draw opacity=0.8, thick]
(axis cs:-0.449173629283905,1)
--(axis cs:0.0415754467248917,1);

\path [draw=black, draw opacity=0.8, thick]
(axis cs:-0.0862921327352524,2)
--(axis cs:0.150536134839058,2);

\path [draw=black, draw opacity=0.8, thick]
(axis cs:-0.454275846481323,3)
--(axis cs:0.015927791595459,3);

\path [draw=black, draw opacity=0.8, thick]
(axis cs:0.16183041036129,4)
--(axis cs:0.505303680896759,4);

\addplot [thick, steelblue52138189, dashed]
table {%
0 0
0 1
0 2
0 3
0 4
};
\addplot [thick, black, opacity=0.8, mark=diamond*, mark size=2.5, mark options={solid}, only marks]
table {%
-0.0248636547476053 0
-0.233725234866142 1
0.0528422445058823 2
-0.21710991859436 3
0.341891795396805 4
};
\end{axis}

\end{tikzpicture}
    }
    \caption{The results of the coefficients $\beta$ for the different Fields of study when using Google Scholar data (left), Semantic Scholar (middle), and Semantic Scholar with the addition of the medical papers (right). The x-axis is coefficient value and the forest plot shows the estimated value and 95\% credible interval.  }
    \label{fig:sci_forest_plots}
\end{figure}

We begin by examining the conclusions inferred by our model on the three versions of the data, Google Scholar, Semantic Scholar, and Semantic Scholar with the medical domain added.  The results can be found in \autoref{fig:sci_forest_plots}, showing consistent conclusions of no correlation between field and citation rate of reproducible papers for any of the three original fields. When Medicine is added we observe that it does show high citation rate for reproducing papers, without changing the conclusion of the other fields. 

Beyond the consistency of the conclusions, we are further confident in our approach's conclusions due to better model fit. The Google Scholar case producing an $R^2 = 0.41$, and the Semantic Scholar data with/without the Medicine papers at $R^2=0.24$ and $0.19$ respectively. We arguably would not expect very high $R^2$ values considering the model is characterizing populations of citation rates based only on the field, as prior work focusing on predicting citations using venue, author, and content information achieved  $R^2 = 0.74$ \citep{10.1145/2063576.2063757}. 

This approach has also provided further insight into the nature of reproduction and citations, that the reward behaviors are not consistent across fields (subject to unobserved confounders). The question then becomes:
do reproducible papers have a different \textit{style} of citation patterns (i.e., accumulating or decreasing citation rates at a different pace) compared to no reproducible work?

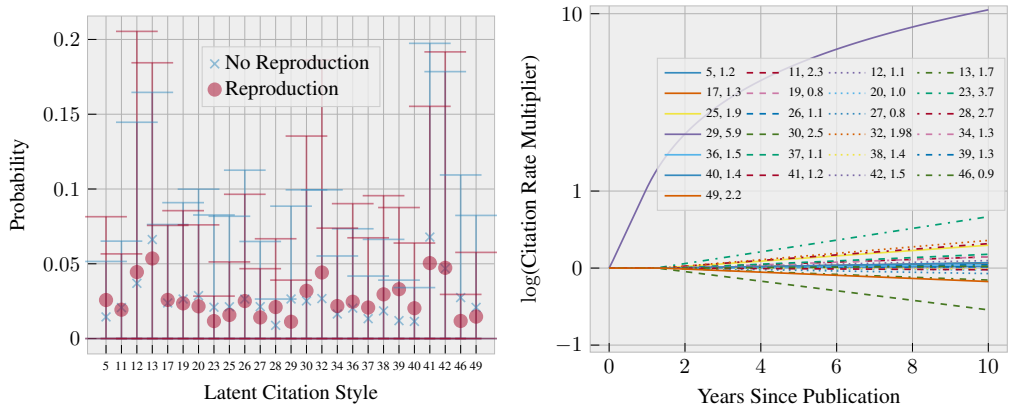
\begin{figure}[!h]
    \centering
    \adjustbox{max width=0.48\columnwidth}{%
    \begin{tikzpicture}

\definecolor{darkgray178}{RGB}{178,178,178}
\definecolor{firebrick166640}{RGB}{166,6,40}
\definecolor{lightgray204}{RGB}{204,204,204}
\definecolor{silver188}{RGB}{188,188,188}
\definecolor{steelblue52138189}{RGB}{52,138,189}
\definecolor{whitesmoke238}{RGB}{238,238,238}

\begin{axis}[
axis background/.style={fill=whitesmoke238},
axis line style={silver188},
legend cell align={left},
legend style={
  fill opacity=0.8,
  draw opacity=1,
  text opacity=1,
  at={(0.5,0.91)},
  anchor=north,
  draw=lightgray204,
  fill=whitesmoke238
},
tick pos=left,
x grid style={darkgray178},
xlabel={Latent Citation Style},
xmajorgrids,
xmin=-1.2, xmax=25.2,
xtick style={color=black},
xtick={0,1,2,3,4,5,6,7,8,9,10,11,12,13,14,15,16,17,18,19,20,21,22,23,24},
xticklabels={5,11,12,13,17,19,20,23,25,26,27,28,29,30,32,34,36,37,38,39,40,41,42,46,49},
xticklabel style = {font=\tiny},
y grid style={darkgray178},
ylabel={Probability},
yticklabel style={
    /pgf/number format/fixed,
    /pgf/number format/precision=2
},
ymajorgrids,
ymin=-0.0102699056267738, ymax=0.215668018162251,
ytick style={color=black}
]
\path [draw=steelblue52138189, draw opacity=0.5, thick]
(axis cs:0,0)
--(axis cs:0,0.0515680015087128);

\path [draw=steelblue52138189, draw opacity=0.5, thick]
(axis cs:1,0)
--(axis cs:1,0.0651421472430229);

\path [draw=steelblue52138189, draw opacity=0.5, thick]
(axis cs:2,0)
--(axis cs:2,0.144627213478088);

\path [draw=steelblue52138189, draw opacity=0.5, thick]
(axis cs:3,0)
--(axis cs:3,0.164603248238564);

\path [draw=steelblue52138189, draw opacity=0.5, thick]
(axis cs:4,0)
--(axis cs:4,0.0762642025947571);

\path [draw=steelblue52138189, draw opacity=0.5, thick]
(axis cs:5,0)
--(axis cs:5,0.0908762365579605);

\path [draw=steelblue52138189, draw opacity=0.5, thick]
(axis cs:6,0)
--(axis cs:6,0.0998886674642563);

\path [draw=steelblue52138189, draw opacity=0.5, thick]
(axis cs:7,0)
--(axis cs:7,0.0826242491602898);

\path [draw=steelblue52138189, draw opacity=0.5, thick]
(axis cs:8,0)
--(axis cs:8,0.0817811638116837);

\path [draw=steelblue52138189, draw opacity=0.5, thick]
(axis cs:9,0)
--(axis cs:9,0.1125822737813);

\path [draw=steelblue52138189, draw opacity=0.5, thick]
(axis cs:10,0)
--(axis cs:10,0.0649125576019287);

\path [draw=steelblue52138189, draw opacity=0.5, thick]
(axis cs:11,0)
--(axis cs:11,0.0264215879142284);

\path [draw=steelblue52138189, draw opacity=0.5, thick]
(axis cs:12,0)
--(axis cs:12,0.0885248407721519);

\path [draw=steelblue52138189, draw opacity=0.5, thick]
(axis cs:13,0)
--(axis cs:13,0.0994310304522514);

\path [draw=steelblue52138189, draw opacity=0.5, thick]
(axis cs:14,0)
--(axis cs:14,0.0994394719600677);

\path [draw=steelblue52138189, draw opacity=0.5, thick]
(axis cs:15,0)
--(axis cs:15,0.0552136227488518);

\path [draw=steelblue52138189, draw opacity=0.5, thick]
(axis cs:16,0)
--(axis cs:16,0.0734159499406815);

\path [draw=steelblue52138189, draw opacity=0.5, thick]
(axis cs:17,0)
--(axis cs:17,0.041793916374445);

\path [draw=steelblue52138189, draw opacity=0.5, thick]
(axis cs:18,0)
--(axis cs:18,0.066308781504631);

\path [draw=steelblue52138189, draw opacity=0.5, thick]
(axis cs:19,0)
--(axis cs:19,0.0390703640878201);

\path [draw=steelblue52138189, draw opacity=0.5, thick]
(axis cs:20,0)
--(axis cs:20,0.0340806059539318);

\path [draw=steelblue52138189, draw opacity=0.5, thick]
(axis cs:21,0)
--(axis cs:21,0.197388887405396);

\path [draw=steelblue52138189, draw opacity=0.5, thick]
(axis cs:22,0)
--(axis cs:22,0.178487032651901);

\path [draw=steelblue52138189, draw opacity=0.5, thick]
(axis cs:23,0)
--(axis cs:23,0.109384156763554);

\path [draw=steelblue52138189, draw opacity=0.5, thick]
(axis cs:24,0)
--(axis cs:24,0.0823396667838097);

\path [draw=firebrick166640, draw opacity=0.5, thick]
(axis cs:0,0)
--(axis cs:0,0.0814763978123665);

\path [draw=firebrick166640, draw opacity=0.5, thick]
(axis cs:1,0)
--(axis cs:1,0.0566180981695652);

\path [draw=firebrick166640, draw opacity=0.5, thick]
(axis cs:2,0)
--(axis cs:2,0.205398112535477);

\path [draw=firebrick166640, draw opacity=0.5, thick]
(axis cs:3,0)
--(axis cs:3,0.184408143162727);

\path [draw=firebrick166640, draw opacity=0.5, thick]
(axis cs:4,0)
--(axis cs:4,0.0756504908204079);

\path [draw=firebrick166640, draw opacity=0.5, thick]
(axis cs:5,0)
--(axis cs:5,0.0855218544602394);

\path [draw=firebrick166640, draw opacity=0.5, thick]
(axis cs:6,0)
--(axis cs:6,0.0760268494486809);

\path [draw=firebrick166640, draw opacity=0.5, thick]
(axis cs:7,0)
--(axis cs:7,0.0283820908516645);

\path [draw=firebrick166640, draw opacity=0.5, thick]
(axis cs:8,0)
--(axis cs:8,0.051255390048027);

\path [draw=firebrick166640, draw opacity=0.5, thick]
(axis cs:9,0)
--(axis cs:9,0.0964687317609787);

\path [draw=firebrick166640, draw opacity=0.5, thick]
(axis cs:10,0)
--(axis cs:10,0.0467267148196697);

\path [draw=firebrick166640, draw opacity=0.5, thick]
(axis cs:11,0)
--(axis cs:11,0.066740095615387);

\path [draw=firebrick166640, draw opacity=0.5, thick]
(axis cs:12,0)
--(axis cs:12,0.0390930734574795);

\path [draw=firebrick166640, draw opacity=0.5, thick]
(axis cs:13,0)
--(axis cs:13,0.13539756834507);

\path [draw=firebrick166640, draw opacity=0.5, thick]
(axis cs:14,0)
--(axis cs:14,0.186201974749565);

\path [draw=firebrick166640, draw opacity=0.5, thick]
(axis cs:15,0)
--(axis cs:15,0.0738432705402374);

\path [draw=firebrick166640, draw opacity=0.5, thick]
(axis cs:16,0)
--(axis cs:16,0.0902154147624969);

\path [draw=firebrick166640, draw opacity=0.5, thick]
(axis cs:17,0)
--(axis cs:17,0.0673748776316643);

\path [draw=firebrick166640, draw opacity=0.5, thick]
(axis cs:18,0)
--(axis cs:18,0.0955386608839035);

\path [draw=firebrick166640, draw opacity=0.5, thick]
(axis cs:19,0)
--(axis cs:19,0.0876086801290512);

\path [draw=firebrick166640, draw opacity=0.5, thick]
(axis cs:20,0)
--(axis cs:20,0.0638743191957474);

\path [draw=firebrick166640, draw opacity=0.5, thick]
(axis cs:21,0)
--(axis cs:21,0.155277878046036);

\path [draw=firebrick166640, draw opacity=0.5, thick]
(axis cs:22,0)
--(axis cs:22,0.191679954528809);

\path [draw=firebrick166640, draw opacity=0.5, thick]
(axis cs:23,0)
--(axis cs:23,0.0295204315334558);

\path [draw=firebrick166640, draw opacity=0.5, thick]
(axis cs:24,0)
--(axis cs:24,0.0576133504509926);

\addplot [thick, steelblue52138189, opacity=0.5, mark=x, mark size=3, mark options={solid}, only marks]
table {%
0 0.0144743351265788
1 0.0207254085689783
2 0.0369492843747139
3 0.06619843095541
4 0.0238094143569469
5 0.0264204069972038
6 0.0286973491311073
7 0.0209673512727022
8 0.0212486907839775
9 0.0264125037938356
10 0.0213342569768429
11 0.00884018372744322
12 0.0264653693884611
13 0.0251681283116341
14 0.0267271734774113
15 0.0164320040494204
16 0.0203704368323088
17 0.0132916159927845
18 0.0184943303465843
19 0.0119786653667688
20 0.0114762270823121
21 0.0678640976548195
22 0.0468351505696774
23 0.0274108946323395
24 0.0208413768559694
};
\addlegendentry{No Reproduction}
\addplot [thick, firebrick166640, opacity=0.5, mark=*, mark size=3, mark options={solid}, only marks]
table {%
0 0.0257682744413614
1 0.0193004570901394
2 0.0444693006575108
3 0.0535499155521393
4 0.0258455462753773
5 0.0235231406986713
6 0.0215582735836506
7 0.0116921868175268
8 0.0157614257186651
9 0.0250720363110304
10 0.0141267077997327
11 0.0210424866527319
12 0.0112514989450574
13 0.0319400019943714
14 0.0441646166145802
15 0.0217463281005621
16 0.0246255956590176
17 0.0207422394305468
18 0.0295604951679707
19 0.0330057926476002
20 0.0202875938266516
21 0.0504497028887272
22 0.0472967363893986
23 0.0117822289466858
24 0.0146691044792533
};
\addlegendentry{Reproduction}
\addplot [thick, steelblue52138189, opacity=0.5, mark=-, mark size=10, mark options={solid}, only marks]
table {%
0 0
1 0
2 0
3 0
4 0
5 0
6 0
7 0
8 0
9 0
10 0
11 0
12 0
13 0
14 0
15 0
16 0
17 0
18 0
19 0
20 0
21 0
22 0
23 0
24 0
};
\addplot [thick, steelblue52138189, opacity=0.5, mark=-, mark size=10, mark options={solid}, only marks]
table {%
0 0.0515680015087128
1 0.0651421472430229
2 0.144627213478088
3 0.164603248238564
4 0.0762642025947571
5 0.0908762365579605
6 0.0998886674642563
7 0.0826242491602898
8 0.0817811638116837
9 0.1125822737813
10 0.0649125576019287
11 0.0264215879142284
12 0.0885248407721519
13 0.0994310304522514
14 0.0994394719600677
15 0.0552136227488518
16 0.0734159499406815
17 0.041793916374445
18 0.066308781504631
19 0.0390703640878201
20 0.0340806059539318
21 0.197388887405396
22 0.178487032651901
23 0.109384156763554
24 0.0823396667838097
};
\addplot [thick, firebrick166640, opacity=0.5, mark=-, mark size=10, mark options={solid}, only marks]
table {%
0 0
1 0
2 0
3 0
4 0
5 0
6 0
7 0
8 0
9 0
10 0
11 0
12 0
13 0
14 0
15 0
16 0
17 0
18 0
19 0
20 0
21 0
22 0
23 0
24 0
};
\addplot [thick, firebrick166640, opacity=0.5, mark=-, mark size=10, mark options={solid}, only marks]
table {%
0 0.0814763978123665
1 0.0566180981695652
2 0.205398112535477
3 0.184408143162727
4 0.0756504908204079
5 0.0855218544602394
6 0.0760268494486809
7 0.0283820908516645
8 0.051255390048027
9 0.0964687317609787
10 0.0467267148196697
11 0.066740095615387
12 0.0390930734574795
13 0.13539756834507
14 0.186201974749565
15 0.0738432705402374
16 0.0902154147624969
17 0.0673748776316643
18 0.0955386608839035
19 0.0876086801290512
20 0.0638743191957474
21 0.155277878046036
22 0.191679954528809
23 0.0295204315334558
24 0.0576133504509926
};

\end{axis}

\end{tikzpicture}
    }
    \adjustbox{max width=0.48\columnwidth}{%
    \input{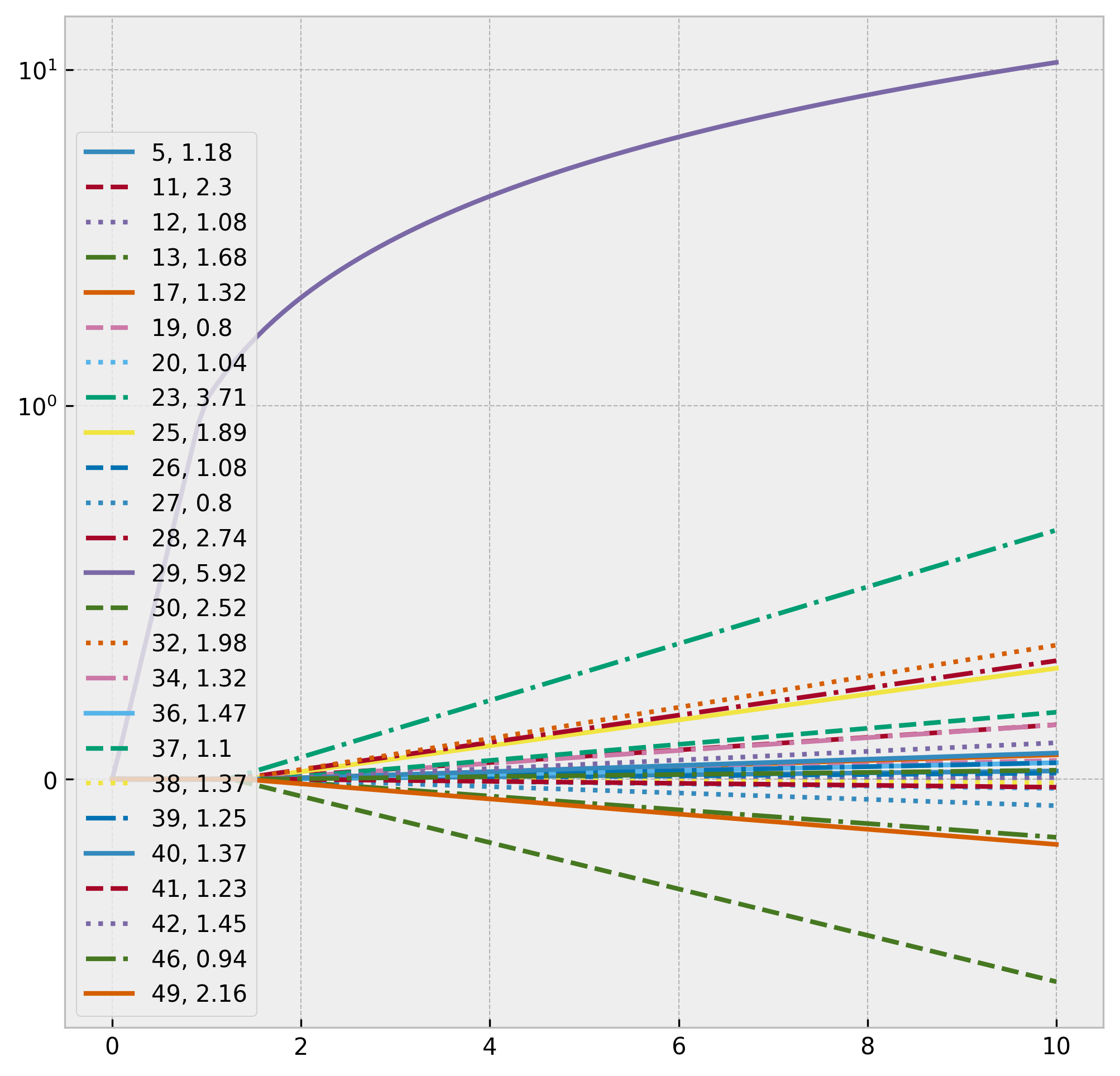}
    }
    \caption{The discovered latent citation styles and their proportion of use in reproduced and failed to reproduce papers (left) and the log multiplicative effect of the citation rate over time (right). Note the right legend shows ``Citation Style, Mean Occurrence Rate of the Style''. The y-axis is a symmetric logarithm scale with linear behavior in the $[-1, 1]$ range, where $0$ indicates no change in the citation rate. Error bars are the 95\% credible interval inferred by the model. }
    \label{fig:sci_style_shapes}
\end{figure}

Per the design of our model, in \autoref{fig:sci_style_shapes} we can investigate the citation rates over time as inferred by our model, shown for the Semantic Scholar + Medicine case. In this instance we do not observe any difference in the citation rates or style between (non)reproducing papers. A maximum of 50 components were allowed for computational tractability, and non-present components are ones the model learned to discard with near-zero probabilities. We note of particular interest that most latent citation styles only have an impact starting two years out from publication, a result consistent with prior work which found the first two years of citations to be highly predictive of the long-term cumulative number of citations \citep{Stegehuis2015}. This provides another degree of confidence in the validity of our general approach, though we do not make claim that our simple model of citation rate is the best possible choice. 

The data is also interesting in that we observe behaviors not normally discussed in bibliometric literature: papers who's citation rate decreases with time. This is indeed not directly observable in the common modeling approach of looking at cumulative citations after a point in time. We further find citation style 29 uniquely interesting as a ``runaway success'', quickly multiplying the citation rate by $\exp(10) \approx 10^{4.35}$ after ten years.

\subsection{Machine Learning Results}

Having shown our model allows for more robust conclusions around the impact replicatiable results has on citation rate, we turn to the machine learning reproductions documented by \cite{Raff2019_quantify_repro}. Many of the papers were selected by the author's personal interests, rather than impact factor, so we do not find it appropriate to include it in the same hierarchical model. The ML data also includes numerous other quantification's about the paper not present in the prior section, so we treat it separately. We use the same approach without a hierarchical prior over field since it is one population of papers. The adjusted $R^2$ of the model is 0.31 using Semantic Scholar for the citation data, inline with the prior experiments. 

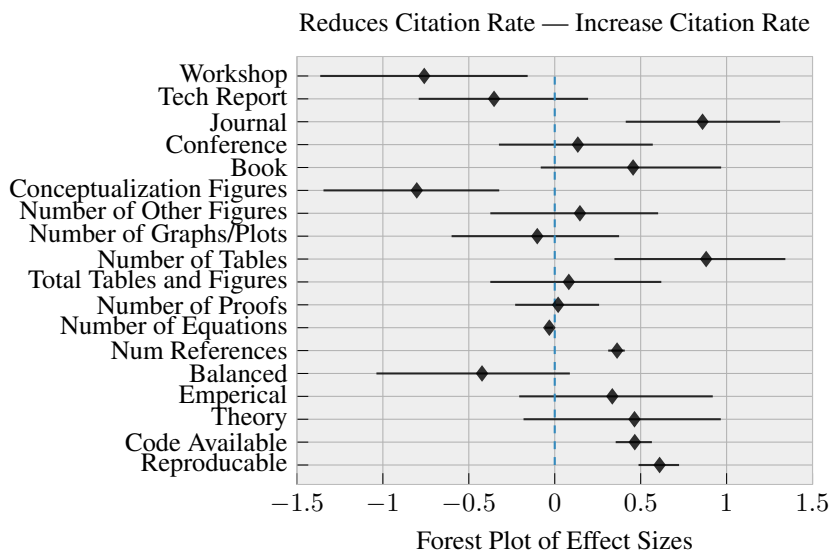
\begin{figure}[!h]
    \centering
    \begin{tikzpicture}

\definecolor{darkgray178}{RGB}{178,178,178}
\definecolor{silver188}{RGB}{188,188,188}
\definecolor{steelblue52138189}{RGB}{52,138,189}
\definecolor{whitesmoke238}{RGB}{238,238,238}

\begin{axis}[
axis background/.style={fill=whitesmoke238},
axis line style={silver188},
tick pos=left,
title={Reduces Citation Rate  | Increase Citation Rate},
x grid style={darkgray178},
xlabel={Forest Plot of Effect Sizes},
xmajorgrids,
xmin=-1.49976090192795, xmax=1.49976090192795,
xtick style={color=black},
y grid style={darkgray178},
ymajorgrids,
ymin=-0.85, ymax=17.85,
ytick style={color=black},
ytick={0,1,2,3,4,5,6,7,8,9,10,11,12,13,14,15,16,17},
yticklabels={
  Reproducable,
  Code Available,
  Theory,
  Emperical,
  Balanced,
  Num References,
  Number of Equations,
  Number of Proofs,
  Total Tables and Figures,
  Number of Tables,
  Number of Graphs/Plots,
  Number of Other Figures,
  Conceptualization Figures,
  Book,
  Conference,
  Journal,
  Tech Report,
  Workshop
}
]
\path [draw=black, draw opacity=0.8, thick]
(axis cs:0.487510710954666,0)
--(axis cs:0.723459303379059,0);

\path [draw=black, draw opacity=0.8, thick]
(axis cs:0.35454136133194,1)
--(axis cs:0.565206170082092,1);

\path [draw=black, draw opacity=0.8, thick]
(axis cs:-0.181585967540741,2)
--(axis cs:0.966238141059875,2);

\path [draw=black, draw opacity=0.8, thick]
(axis cs:-0.206831187009811,3)
--(axis cs:0.919521808624268,3);

\path [draw=black, draw opacity=0.8, thick]
(axis cs:-1.03752148151398,4)
--(axis cs:0.0879292786121368,4);

\path [draw=black, draw opacity=0.8, thick]
(axis cs:0.311009526252747,5)
--(axis cs:0.407429367303848,5);

\path [draw=black, draw opacity=0.8, thick]
(axis cs:-0.0565976686775684,6)
--(axis cs:-0.0104642994701862,6);

\path [draw=black, draw opacity=0.8, thick]
(axis cs:-0.230207830667496,7)
--(axis cs:0.258203417062759,7);

\path [draw=black, draw opacity=0.8, thick]
(axis cs:-0.375093251466751,8)
--(axis cs:0.62000834941864,8);

\path [draw=black, draw opacity=0.8, thick]
(axis cs:0.347488760948181,9)
--(axis cs:1.34183859825134,9);

\path [draw=black, draw opacity=0.8, thick]
(axis cs:-0.600265502929688,10)
--(axis cs:0.374452531337738,10);

\path [draw=black, draw opacity=0.8, thick]
(axis cs:-0.374829232692719,11)
--(axis cs:0.601554691791534,11);

\path [draw=black, draw opacity=0.8, thick]
(axis cs:-1.34553110599518,12)
--(axis cs:-0.323083192110062,12);

\path [draw=black, draw opacity=0.8, thick]
(axis cs:-0.0814915299415588,13)
--(axis cs:0.968190371990204,13);

\path [draw=black, draw opacity=0.8, thick]
(axis cs:-0.324816226959229,14)
--(axis cs:0.570476472377777,14);

\path [draw=black, draw opacity=0.8, thick]
(axis cs:0.413084656000137,15)
--(axis cs:1.3110294342041,15);

\path [draw=black, draw opacity=0.8, thick]
(axis cs:-0.791465580463409,16)
--(axis cs:0.193963319063187,16);

\path [draw=black, draw opacity=0.8, thick]
(axis cs:-1.36444664001465,17)
--(axis cs:-0.157528281211853,17);

\addplot [thick, steelblue52138189, dashed]
table {%
0 0
0 1
0 2
0 3
0 4
0 5
0 6
0 7
0 8
0 9
0 10
0 11
0 12
0 13
0 14
0 15
0 16
0 17
};
\addplot [thick, black, opacity=0.8, mark=diamond*, mark size=2.5, mark options={solid}, only marks]
table {%
0.610106587409973 0
0.465711802244186 1
0.464042067527771 2
0.334863573312759 3
-0.42306461930275 4
0.362679064273834 5
-0.0315931476652622 6
0.019747830927372 7
0.0819313228130341 8
0.881074070930481 9
-0.101658925414085 10
0.146564200520515 11
-0.802769064903259 12
0.455967247486115 13
0.134346574544907 14
0.860010981559753 15
-0.35307189822197 16
-0.759182929992676 17
};
\end{axis}

\end{tikzpicture}
    \caption{Forest plot of the coefficients $\beta$ of various features, with 95\% credible intervals. }
    \label{fig:ml_forest}
\end{figure}

The results \autoref{fig:ml_forest} show that reproducible papers, and papers that make their code available, both receive higher rates of citation. The former is desirable, and the later indicates a strong motivation for authors to open source their code beyond the arguments around replication~\citep{soton403913,Claerbout1992,Callahan2016,10.7554/eLife.67995,Forde2018}. The sharing of code is generally argued to be beneficial, but we do note that it captures methodological flaws as well --- and is thus not a panacea to concerns around reproduction \citep{dror-etal-2019-deep,Dror2017,dror-etal-2018-hitchhikers,10.1145/3383313.3412489,pmlr-v97-bouthillier19a,Bouthillier2021}. We are also encouraged that more references per page has a higher citation rate, under the belief this corresponds to more thorough documentation of prior work and good scholastic behaviors. 

The reduced citation rate for Conceptualization Figures, which attempt to convey the intuition of a method, is interesting. \citet{Raff2019_quantify_repro} noted no relationship between this variable and replication, while later work found that papers which use conceptualization figures take less time/human effort to reproduce ~\citep{Raff2020c}. This type of scientific communication appears to have a particularly complex relationship with reproduction and the incentives around reproduction that thus warrants further study. 

The last points of note are that publishing in Journals, and more tables appear to increase citation rate while publishing in a workshop reduces it. Publishing in a workshop having a lower citation rate makes sense intuitively, though it is perhaps interesting that tech reports (like arXiv)  have no relationship --- and it is worth studying whether workshops being a final ``home'' for a paper may have a negative perception. This result is also possible due to the noted bias in the data, which we believe may explain the result that Journal publications have a higher citation rate, since ML as a field generally prefers conferences over journals. Last, we have no particular intuition about why having more tables per paper may lead to more citations --- unless it is a matter of making it easy for future papers to re-use the reported results, a hypothesis proposed in \citep{Raff2019_quantify_repro}.

\begin{figure}[!h]
    \centering
    \adjustbox{max width=0.48\columnwidth}{%
    \begin{tikzpicture}

\definecolor{darkgray178}{RGB}{178,178,178}
\definecolor{firebrick166640}{RGB}{166,6,40}
\definecolor{lightgray204}{RGB}{204,204,204}
\definecolor{silver188}{RGB}{188,188,188}
\definecolor{steelblue52138189}{RGB}{52,138,189}
\definecolor{whitesmoke238}{RGB}{238,238,238}

\begin{axis}[
axis background/.style={fill=whitesmoke238},
axis line style={silver188},
legend cell align={left},
legend style={fill opacity=0.8, draw opacity=1, text opacity=1, draw=lightgray204, fill=whitesmoke238},
tick pos=left,
x grid style={darkgray178},
xlabel={Latent Citation Style},
xmajorgrids,
xmin=-0.7, xmax=14.7,
xtick style={color=black},
xtick={0,1,2,3,4,5,6,7,8,9,10,11,12,13,14},
xticklabels={1,3,10,11,13,16,18,23,28,29,30,32,35,44,49},
y grid style={darkgray178},
ylabel={Probability},
yticklabel style={
    /pgf/number format/fixed,
    /pgf/number format/precision=2
},
ymajorgrids,
ymin=-0.0119153007864952, ymax=0.250221316516399,
ytick style={color=black}
]
\path [draw=steelblue52138189, draw opacity=0.5, thick]
(axis cs:0,0)
--(axis cs:0,0.0757121965289116);

\path [draw=steelblue52138189, draw opacity=0.5, thick]
(axis cs:1,0)
--(axis cs:1,0.238306015729904);

\path [draw=steelblue52138189, draw opacity=0.5, thick]
(axis cs:2,0)
--(axis cs:2,0.100814715027809);

\path [draw=steelblue52138189, draw opacity=0.5, thick]
(axis cs:3,0)
--(axis cs:3,0.102215811610222);

\path [draw=steelblue52138189, draw opacity=0.5, thick]
(axis cs:4,0)
--(axis cs:4,0.111107721924782);

\path [draw=steelblue52138189, draw opacity=0.5, thick]
(axis cs:5,0)
--(axis cs:5,0.205615654587746);

\path [draw=steelblue52138189, draw opacity=0.5, thick]
(axis cs:6,0.0309693738818169)
--(axis cs:6,0.126351550221443);

\path [draw=steelblue52138189, draw opacity=0.5, thick]
(axis cs:7,0)
--(axis cs:7,0.165140315890312);

\path [draw=steelblue52138189, draw opacity=0.5, thick]
(axis cs:8,0)
--(axis cs:8,0.0723772644996643);

\path [draw=steelblue52138189, draw opacity=0.5, thick]
(axis cs:9,0)
--(axis cs:9,0.0104431081563234);

\path [draw=steelblue52138189, draw opacity=0.5, thick]
(axis cs:10,0)
--(axis cs:10,0.0755969062447548);

\path [draw=steelblue52138189, draw opacity=0.5, thick]
(axis cs:11,0)
--(axis cs:11,0.0413766615092754);

\path [draw=steelblue52138189, draw opacity=0.5, thick]
(axis cs:12,0)
--(axis cs:12,0.21881602704525);

\path [draw=steelblue52138189, draw opacity=0.5, thick]
(axis cs:13,0)
--(axis cs:13,0.0913147926330566);

\path [draw=steelblue52138189, draw opacity=0.5, thick]
(axis cs:14,0)
--(axis cs:14,0.0990901589393616);

\path [draw=firebrick166640, draw opacity=0.5, thick]
(axis cs:0,0)
--(axis cs:0,0.0656364932656288);

\path [draw=firebrick166640, draw opacity=0.5, thick]
(axis cs:1,0)
--(axis cs:1,0.134890645742416);

\path [draw=firebrick166640, draw opacity=0.5, thick]
(axis cs:2,0)
--(axis cs:2,0.0732557624578476);

\path [draw=firebrick166640, draw opacity=0.5, thick]
(axis cs:3,1.67638063430786e-08)
--(axis cs:3,0.0792496576905251);

\path [draw=firebrick166640, draw opacity=0.5, thick]
(axis cs:4,0)
--(axis cs:4,0.0393326580524445);

\path [draw=firebrick166640, draw opacity=0.5, thick]
(axis cs:5,0)
--(axis cs:5,0.16072216629982);

\path [draw=firebrick166640, draw opacity=0.5, thick]
(axis cs:6,0.0247739218175411)
--(axis cs:6,0.0912539660930634);

\path [draw=firebrick166640, draw opacity=0.5, thick]
(axis cs:7,0)
--(axis cs:7,0.114678397774696);

\path [draw=firebrick166640, draw opacity=0.5, thick]
(axis cs:8,0)
--(axis cs:8,0.11752313375473);

\path [draw=firebrick166640, draw opacity=0.5, thick]
(axis cs:9,0.0559124015271664)
--(axis cs:9,0.124144501984119);

\path [draw=firebrick166640, draw opacity=0.5, thick]
(axis cs:10,0)
--(axis cs:10,0.0845465883612633);

\path [draw=firebrick166640, draw opacity=0.5, thick]
(axis cs:11,0)
--(axis cs:11,0.0928247570991516);

\path [draw=firebrick166640, draw opacity=0.5, thick]
(axis cs:12,0)
--(axis cs:12,0.150442525744438);

\path [draw=firebrick166640, draw opacity=0.5, thick]
(axis cs:13,0)
--(axis cs:13,0.10518392175436);

\path [draw=firebrick166640, draw opacity=0.5, thick]
(axis cs:14,0)
--(axis cs:14,0.0722800120711327);

\addplot [thick, steelblue52138189, opacity=0.5, mark=x, mark size=3, mark options={solid}, only marks]
table {%
0 0.0208419766277075
1 0.0776373967528343
2 0.025570971891284
3 0.0286173354834318
4 0.0262917410582304
5 0.060882568359375
6 0.0799310505390167
7 0.0447162166237831
8 0.0213250573724508
9 0.00288877496495843
10 0.023232152685523
11 0.0145326191559434
12 0.084339439868927
13 0.0460446178913116
14 0.0250144172459841
};
\addlegendentry{No Reproduction}
\addplot [thick, firebrick166640, opacity=0.5, mark=*, mark size=3, mark options={solid}, only marks]
table {%
0 0.018362395465374
1 0.04605358466506
2 0.0202278569340706
3 0.0198853611946106
4 0.0114284176379442
5 0.0464632362127304
6 0.0578494407236576
7 0.0327972806990147
8 0.0307484529912472
9 0.0908110439777374
10 0.0263623166829348
11 0.0227774512022734
12 0.0616738572716713
13 0.0635411515831947
14 0.0197501890361309
};
\addlegendentry{Reproduction}
\addplot [thick, steelblue52138189, opacity=0.5, mark=-, mark size=10, mark options={solid}, only marks]
table {%
0 0
1 0
2 0
3 0
4 0
5 0
6 0.0309693738818169
7 0
8 0
9 0
10 0
11 0
12 0
13 0
14 0
};
\addplot [thick, steelblue52138189, opacity=0.5, mark=-, mark size=10, mark options={solid}, only marks]
table {%
0 0.0757121965289116
1 0.238306015729904
2 0.100814715027809
3 0.102215811610222
4 0.111107721924782
5 0.205615654587746
6 0.126351550221443
7 0.165140315890312
8 0.0723772644996643
9 0.0104431081563234
10 0.0755969062447548
11 0.0413766615092754
12 0.21881602704525
13 0.0913147926330566
14 0.0990901589393616
};
\addplot [thick, firebrick166640, opacity=0.5, mark=-, mark size=10, mark options={solid}, only marks]
table {%
0 0
1 0
2 0
3 1.67638063430786e-08
4 0
5 0
6 0.0247739218175411
7 0
8 0
9 0.0559124015271664
10 0
11 0
12 0
13 0
14 0
};
\addplot [thick, firebrick166640, opacity=0.5, mark=-, mark size=10, mark options={solid}, only marks]
table {%
0 0.0656364932656288
1 0.134890645742416
2 0.0732557624578476
3 0.0792496576905251
4 0.0393326580524445
5 0.16072216629982
6 0.0912539660930634
7 0.114678397774696
8 0.11752313375473
9 0.124144501984119
10 0.0845465883612633
11 0.0928247570991516
12 0.150442525744438
13 0.10518392175436
14 0.0722800120711327
};

\end{axis}

\end{tikzpicture}
    }
    \adjustbox{max width=0.48\columnwidth}{%
    \input{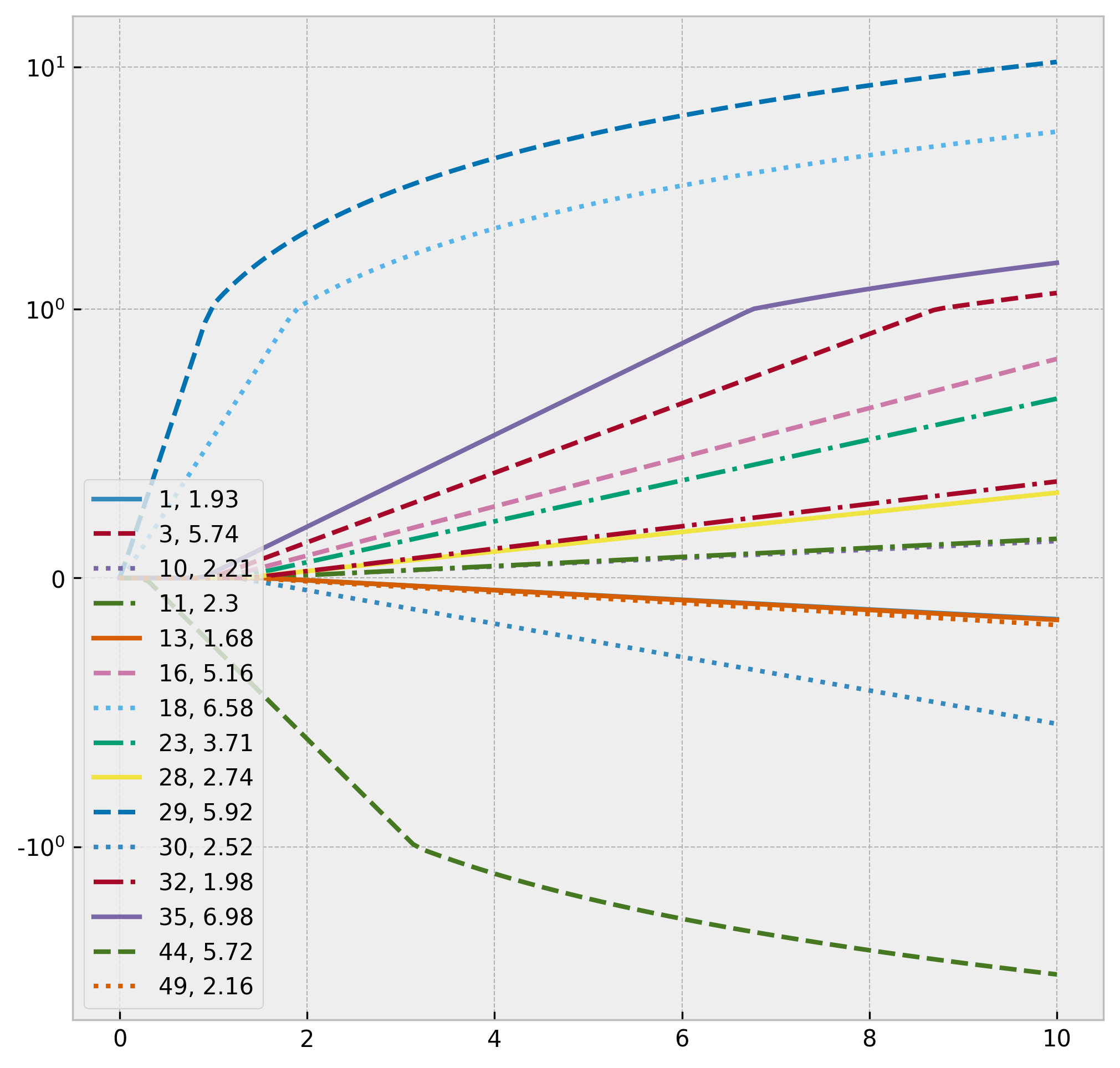}
    }
    \caption{The discovered latent citation styles and their proportion of use in reproduced and failed to reproduce papers (left) and the log multiplicative effect of the citation rate over time (right) for the Machine Learning data. Note the right legend shows ``Citation Style, Mean Occurrence Rate of the Style''. The y-axis is a symmetric logarithm scale with linear behavior in the $[-1, 1]$ range, where $0$ indicates no change in the citation rate.}
    \label{fig:ml_style_shapes}
\end{figure}
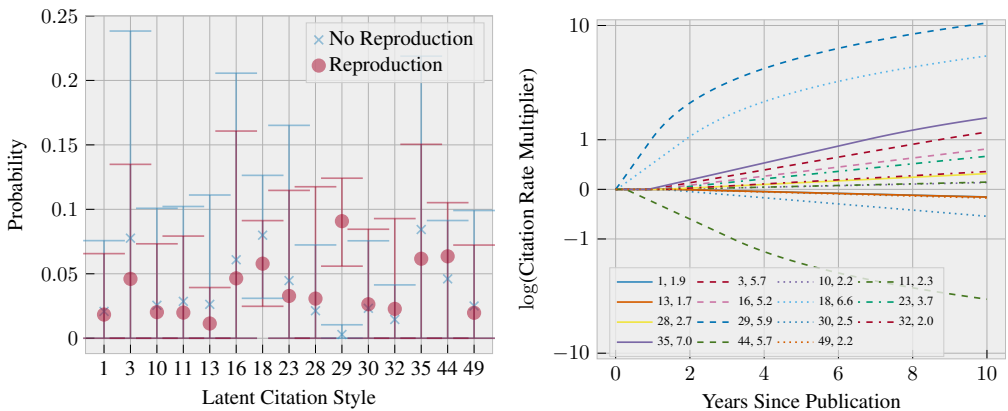

Last, we look at the latent citation patterns again in \autoref{fig:ml_style_shapes}, and note that style 29 does has a significant difference between reproducible and non-reproducible works\footnote{This is a different style ``29'' than in the prior section, it is by pure chance that it also happened to be the 29'th component of the same nature. This took us many hours to ``debug'' to find no apparent but, just random chance.}. By chance this component again represents a ``runaway success'', indicating a preference for degree of success toward reproducible works.

\section{Conclusion} \label{sec:conclusion}

Our results are overall encouraging toward the question of replication and citation: citations are positively correlated or are independent of replication, which is better than the prior hypothesis that non-reproducible works get more citations. Our results for machine learning in particular indicate that citations correlate positively with further desirable behaviors like thorough citations and sharing of code. This work has furthered the bibliometric study of the interaction between citation rate and replication, and we note further valuable directions remain. A large amount of data without ground-truth replication success exists to merge into such analyses, as well as the possibility of using natural language processing to make inferences about paper replications by the content of citing documents. 

\bibliography{references,extra}

\begin{thebibliography}{44}
\providecommand{\natexlab}[1]{#1}
\providecommand{\url}[1]{\texttt{#1}}
\expandafter\ifx\csname urlstyle\endcsname\relax
  \providecommand{\doi}[1]{doi: #1}\else
  \providecommand{\doi}{doi: \begingroup \urlstyle{rm}\Url}\fi

\bibitem[Baker(2016)]{Baker2016}
Monya Baker.
\newblock {1,500 scientists lift the lid on reproducibility}.
\newblock \emph{Nature}, 533\penalty0 (7604):\penalty0 452--454, 2016.
\newblock ISSN 1476-4687.
\newblock \doi{10.1038/533452a}.
\newblock URL \url{https://doi.org/10.1038/533452a}.

\bibitem[Bornmann \& Mutz(2015)Bornmann and Mutz]{Bornmann2015}
Lutz Bornmann and Rüdiger Mutz.
\newblock {Growth rates of modern science: A bibliometric analysis based on the
  number of publications and cited references}.
\newblock \emph{Journal of the Association for Information Science and
  Technology}, 66\penalty0 (11):\penalty0 2215--2222, 2015.
\newblock ISSN 23301643.
\newblock \doi{10.1002/asi.23329}.

\bibitem[Bouthillier et~al.(2019)Bouthillier, Laurent, and
  Vincent]{pmlr-v97-bouthillier19a}
Xavier Bouthillier, César Laurent, and Pascal Vincent.
\newblock {Unreproducible Research is Reproducible}.
\newblock In Kamalika Chaudhuri and Ruslan Salakhutdinov (eds.),
  \emph{Proceedings of the 36th International Conference on Machine Learning},
  volume~97 of \emph{Proceedings of Machine Learning Research}, pp.\  725--734,
  Long Beach, California, USA, 2019. PMLR.
\newblock URL \url{http://proceedings.mlr.press/v97/bouthillier19a.html}.

\bibitem[Bouthillier et~al.(2021)Bouthillier, Delaunay, Bronzi, Trofimov,
  Nichyporuk, Szeto, Sepah, Raff, Madan, Voleti, Kahou, Michalski, Serdyuk,
  Arbel, Pal, Varoquaux, and Vincent]{Bouthillier2021}
Xavier Bouthillier, Pierre Delaunay, Mirko Bronzi, Assya Trofimov, Brennan
  Nichyporuk, Justin Szeto, Naz Sepah, Edward Raff, Kanika Madan, Vikram
  Voleti, Samira~Ebrahimi Kahou, Vincent Michalski, Dmitriy Serdyuk, Tal Arbel,
  Chris Pal, Gaël Varoquaux, and Pascal Vincent.
\newblock {Accounting for Variance in Machine Learning Benchmarks}.
\newblock In \emph{Machine Learning and Systems (MLSys)}, 2021.
\newblock URL \url{http://arxiv.org/abs/2103.03098}.

\bibitem[Callahan et~al.(2016)Callahan, Proctor, Relman, Fukuyama, and
  Holmes]{Callahan2016}
Benjamin Callahan, Diana Proctor, David Relman, Julia Fukuyama, and Susan
  Holmes.
\newblock {REPRODUCIBLE RESEARCH WORKFLOW IN R FOR THE ANALYSIS OF PERSONALIZED
  HUMAN MICROBIOME DATA.}
\newblock \emph{Pacific Symposium on Biocomputing. Pacific Symposium on
  Biocomputing}, 21:\penalty0 183--94, 2016.
\newblock ISSN 2335-6936.
\newblock URL \url{http://www.ncbi.nlm.nih.gov/pubmed/26776185
  http://www.pubmedcentral.nih.gov/articlerender.fcgi?artid=PMC4873301}.

\bibitem[Camerer et~al.(2016)Camerer, Dreber, Forsell, Ho, Huber, Johannesson,
  Kirchler, Almenberg, Altmejd, Chan, Heikensten, Holzmeister, Imai, Isaksson,
  Nave, Pfeiffer, Razen, and Wu]{RePEc:pra:mprapa:75461}
Colin~F. Camerer, Anna Dreber, Eskil Forsell, Teck-Hua Ho, Jürgen Huber,
  Magnus Johannesson, Michael Kirchler, Johan Almenberg, Adam Altmejd, Taizan
  Chan, Emma Heikensten, Felix Holzmeister, Taisuke Imai, Siri Isaksson, Gideon
  Nave, Thomas Pfeiffer, Michael Razen, and Hang Wu.
\newblock {Evaluating replicability of laboratory experiments in economics}.
\newblock \emph{Science}, 351\penalty0 (6280):\penalty0 1433--1436, 3 2016.
\newblock ISSN 0036-8075.
\newblock \doi{10.1126/science.aaf0918}.
\newblock URL \url{https://www.science.org/doi/10.1126/science.aaf0918}.

\bibitem[Camerer et~al.(2018)Camerer, Dreber, Holzmeister, Ho, Huber,
  Johannesson, Kirchler, Nave, Nosek, Pfeiffer, Altmejd, Buttrick, Chan, Chen,
  Forsell, Gampa, Heikensten, Hummer, Imai, Isaksson, Manfredi, Rose,
  Wagenmakers, and Wu]{Camerer2018}
Colin~F. Camerer, Anna Dreber, Felix Holzmeister, Teck~Hua Ho, Jürgen Huber,
  Magnus Johannesson, Michael Kirchler, Gideon Nave, Brian~A. Nosek, Thomas
  Pfeiffer, Adam Altmejd, Nick Buttrick, Taizan Chan, Yiling Chen, Eskil
  Forsell, Anup Gampa, Emma Heikensten, Lily Hummer, Taisuke Imai, Siri
  Isaksson, Dylan Manfredi, Julia Rose, Eric~Jan Wagenmakers, and Hang Wu.
\newblock {Evaluating the replicability of social science experiments in Nature
  and Science between 2010 and 2015}.
\newblock \emph{Nature Human Behaviour}, 2\penalty0 (9):\penalty0 637--644,
  2018.
\newblock ISSN 23973374.
\newblock \doi{10.1038/s41562-018-0399-z}.

\bibitem[Cameron \& Trivedi(1990)Cameron and Trivedi]{Cameron1990}
A.Colin Cameron and Pravin~K Trivedi.
\newblock {Regression-based tests for overdispersion in the Poisson model}.
\newblock \emph{Journal of Econometrics}, 46\penalty0 (3):\penalty0 347--364,
  1990.
\newblock ISSN 0304-4076.
\newblock \doi{https://doi.org/10.1016/0304-4076(90)90014-K}.
\newblock URL
  \url{https://www.sciencedirect.com/science/article/pii/030440769090014K}.

\bibitem[Claerbout \& Karrenbach(1992)Claerbout and Karrenbach]{Claerbout1992}
Jon~F. Claerbout and Martin Karrenbach.
\newblock {Electronic documents give reproducible research a new meaning}.
\newblock In \emph{SEG Technical Program Expanded Abstracts 1992}, pp.\
  601--604. Society of Exploration Geophysicists, 1 1992.
\newblock \doi{10.1190/1.1822162}.
\newblock URL \url{http://library.seg.org/doi/abs/10.1190/1.1822162}.

\bibitem[Collaboration(2015)]{aac4716}
Open~Science Collaboration.
\newblock {Estimating the reproducibility of psychological science}.
\newblock \emph{Science}, 349\penalty0 (6251), 2015.
\newblock ISSN 0036-8075.
\newblock \doi{10.1126/science.aac4716}.
\newblock URL \url{https://science.sciencemag.org/content/349/6251/aac4716}.

\bibitem[Dietz et~al.(2007)Dietz, Bickel, and
  Scheffer]{10.1145/1273496.1273526}
Laura Dietz, Steffen Bickel, and Tobias Scheffer.
\newblock {Unsupervised Prediction of Citation Influences}.
\newblock In \emph{Proceedings of the 24th International Conference on Machine
  Learning}, ICML '07, pp.\  233–240, New York, NY, USA, 2007. Association
  for Computing Machinery.
\newblock ISBN 9781595937933.
\newblock \doi{10.1145/1273496.1273526}.
\newblock URL \url{https://doi.org/10.1145/1273496.1273526}.

\bibitem[Donoho et~al.(2009)Donoho, Maleki, Rahman, Shahram, and
  Stodden]{Donoho2009}
David~L. Donoho, Arian Maleki, Inam~Ur Rahman, Morteza Shahram, and Victoria
  Stodden.
\newblock {Reproducible Research in Computational Harmonic Analysis}.
\newblock \emph{Computing in Science {\&} Engineering}, 11\penalty0
  (1):\penalty0 8--18, 1 2009.
\newblock ISSN 1521-9615.
\newblock \doi{10.1109/MCSE.2009.15}.
\newblock URL \url{http://ieeexplore.ieee.org/document/4720218/}.

\bibitem[Dror et~al.(2017)Dror, Baumer, Bogomolov, and Reichart]{Dror2017}
Rotem Dror, Gili Baumer, Marina Bogomolov, and Roi Reichart.
\newblock {Replicability Analysis for Natural Language Processing: Testing
  Significance with Multiple Datasets}.
\newblock \emph{Transactions of the Association for Computational Linguistics},
  5:\penalty0 471--486, 11 2017.
\newblock ISSN 2307-387X.
\newblock \doi{10.1162/tacl{\_}a{\_}00074}.
\newblock URL \url{https://doi.org/10.1162/tacl_a_00074}.

\bibitem[Dror et~al.(2018)Dror, Baumer, Shlomov, and
  Reichart]{dror-etal-2018-hitchhikers}
Rotem Dror, Gili Baumer, Segev Shlomov, and Roi Reichart.
\newblock {The Hitchhiker's Guide to Testing Statistical Significance in
  Natural Language Processing}.
\newblock In \emph{Proceedings of the 56th Annual Meeting of the Association
  for Computational Linguistics (Volume 1: Long Papers)}, pp.\  1383--1392,
  Melbourne, Australia, 7 2018. Association for Computational Linguistics.
\newblock \doi{10.18653/v1/P18-1128}.
\newblock URL \url{https://aclanthology.org/P18-1128}.

\bibitem[Dror et~al.(2019)Dror, Shlomov, and Reichart]{dror-etal-2019-deep}
Rotem Dror, Segev Shlomov, and Roi Reichart.
\newblock {Deep Dominance - How to Properly Compare Deep Neural Models}.
\newblock In \emph{Proceedings of the 57th Annual Meeting of the Association
  for Computational Linguistics}, pp.\  2773--2785, Florence, Italy, 7 2019.
  Association for Computational Linguistics.
\newblock \doi{10.18653/v1/P19-1266}.
\newblock URL \url{https://aclanthology.org/P19-1266}.

\bibitem[Errington et~al.(2021)Errington, Denis, Perfito, Iorns, and
  Nosek]{10.7554/eLife.67995}
Timothy~M Errington, Alexandria Denis, Nicole Perfito, Elizabeth Iorns, and
  Brian~A Nosek.
\newblock {Reproducibility in Cancer Biology: Challenges for assessing
  replicability in preclinical cancer biology}.
\newblock \emph{eLife}, 10:\penalty0 e67995, 12 2021.
\newblock ISSN 2050-084X.
\newblock \doi{10.7554/eLife.67995}.
\newblock URL \url{https://doi.org/10.7554/eLife.67995}.

\bibitem[Forde et~al.(2018)Forde, Head, Holdgraf, Panda, Perez, Nalvarte,
  Ragan-kelley, and Sundell]{Forde2018}
Jessica Forde, Tim Head, Chris Holdgraf, Yuvi Panda, Fernando Perez, Gladys
  Nalvarte, Benjamin Ragan-kelley, and Erik Sundell.
\newblock {Reproducible Research Environments with repo2docker}.
\newblock In \emph{Reproducibility in ML Workshop, ICML'18}, 2018.

\bibitem[Gardner et~al.(2018)Gardner, Brooks, and Baker]{Gardner2018}
Josh Gardner, Christopher Brooks, and Ryan~S Baker.
\newblock {Enabling End-To-End Machine Learning Replicability : A Case Study in
  Educational Data Mining}.
\newblock In \emph{Reproducibility in ML Workshop, ICML'18}, 2018.

\bibitem[Hoffman \& Gelman(2014)Hoffman and Gelman]{JMLR:v15:hoffman14a}
Matthew~D Hoffman and Andrew Gelman.
\newblock {The No-U-Turn Sampler: Adaptively Setting Path Lengths in
  Hamiltonian Monte Carlo}.
\newblock \emph{Journal of Machine Learning Research}, 15\penalty0
  (47):\penalty0 1593--1623, 2014.
\newblock URL \url{http://jmlr.org/papers/v15/hoffman14a.html}.

\bibitem[Hutson(2018)]{Hutson725}
Matthew Hutson.
\newblock {Artificial intelligence faces reproducibility crisis}.
\newblock \emph{Science}, 359\penalty0 (6377):\penalty0 725--726, 2018.
\newblock ISSN 0036-8075.
\newblock \doi{10.1126/science.359.6377.725}.
\newblock URL \url{https://science.sciencemag.org/content/359/6377/725}.

\bibitem[Kluyver et~al.(2016)Kluyver, Ragan-Kelley, P{\'{e}}rez, Granger,
  Bussonnier, Frederic, Kelley, Hamrick, Grout, Corlay, Ivanov, Avila, Abdalla,
  Willing, and development team]{soton403913}
Thomas Kluyver, Benjamin Ragan-Kelley, Fernando P{\'{e}}rez, Brian Granger,
  Matthias Bussonnier, Jonathan Frederic, Kyle Kelley, Jessica Hamrick, Jason
  Grout, Sylvain Corlay, Paul Ivanov, Damián Avila, Safia Abdalla, Carol
  Willing, and Jupyter development team.
\newblock {Jupyter Notebooks - a publishing format for reproducible
  computational workflows}.
\newblock In Fernando Loizides and Birgit Scmidt (eds.), \emph{Positioning and
  Power in Academic Publishing: Players, Agents and Agendas}, pp.\  87--90. IOS
  Press, 2016.
\newblock URL \url{https://eprints.soton.ac.uk/403913/}.

\bibitem[Kunnath et~al.(2022)Kunnath, Herrmannova, Pride, and
  Knoth]{Kunnath2022}
Suchetha~N Kunnath, Drahomira Herrmannova, David Pride, and Petr Knoth.
\newblock {A meta-analysis of semantic classification of citations}.
\newblock \emph{Quantitative Science Studies}, 2\penalty0 (4):\penalty0
  1170--1215, 2 2022.
\newblock ISSN 2641-3337.
\newblock \doi{10.1162/qss{\_}a{\_}00159}.
\newblock URL \url{https://doi.org/10.1162/qss_a_00159}.

\bibitem[Lotka(1926)]{Lotka1926}
Alfred~J Lotka.
\newblock {The frequency distribution of scientific productivity}.
\newblock \emph{Journal of the Washington Academy of Sciences}, 16\penalty0
  (12):\penalty0 317--323, 10 1926.
\newblock ISSN 00430439.
\newblock URL \url{http://www.jstor.org/stable/24529203}.

\bibitem[Phan et~al.(2019)Phan, Pradhan, and Jankowiak]{Phan2019}
Du~Phan, Neeraj Pradhan, and Martin Jankowiak.
\newblock {Composable Effects for Flexible and Accelerated Probabilistic
  Programming in NumPyro}.
\newblock \emph{arXiv}, pp.\  1--10, 2019.
\newblock URL \url{http://arxiv.org/abs/1912.11554}.

\bibitem[Plesser(2018)]{Plesser2018}
Hans~E Plesser.
\newblock {Reproducibility vs. Replicability: A Brief History of a Confused
  Terminology}.
\newblock \emph{Frontiers in neuroinformatics}, 11:\penalty0 76, 1 2018.
\newblock ISSN 1662-5196.
\newblock \doi{10.3389/fninf.2017.00076}.
\newblock URL \url{https://pubmed.ncbi.nlm.nih.gov/29403370
  https://www.ncbi.nlm.nih.gov/pmc/articles/PMC5778115/}.

\bibitem[Poldrack(2019)]{Poldrack2019}
Russell~A. Poldrack.
\newblock {The Costs of Reproducibility}.
\newblock \emph{Neuron}, 101\penalty0 (1):\penalty0 11--14, 1 2019.
\newblock ISSN 08966273.
\newblock \doi{10.1016/j.neuron.2018.11.030}.
\newblock URL
  \url{https://linkinghub.elsevier.com/retrieve/pii/S0896627318310390}.

\bibitem[Potter(1981)]{Potter1981LotkasLR}
William~Gray Potter.
\newblock {Lotka's Law Revisited}.
\newblock \emph{Library Trends}, 30:\penalty0 21--39, 1981.

\bibitem[Price(1976)]{https://doi.org/10.1002/asi.4630270505}
Derek De~Solla Price.
\newblock {A general theory of bibliometric and other cumulative advantage
  processes}.
\newblock \emph{Journal of the American Society for Information Science},
  27\penalty0 (5):\penalty0 292--306, 1976.
\newblock \doi{https://doi.org/10.1002/asi.4630270505}.
\newblock URL
  \url{https://asistdl.onlinelibrary.wiley.com/doi/abs/10.1002/asi.4630270505}.

\bibitem[Price(1965)]{Price1965}
Derek J. de~Solla Price.
\newblock {Networks of Scientific Papers}.
\newblock \emph{Science}, 149\penalty0 (3683):\penalty0 510--515, 7 1965.
\newblock ISSN 0036-8075.
\newblock \doi{10.1126/science.149.3683.510}.
\newblock URL \url{https://www.science.org/doi/10.1126/science.149.3683.510}.

\bibitem[Raff(2019)]{Raff2019_quantify_repro}
Edward Raff.
\newblock {A Step Toward Quantifying Independently Reproducible Machine
  Learning Research}.
\newblock In \emph{NeurIPS}, 2019.
\newblock URL \url{http://arxiv.org/abs/1909.06674}.

\bibitem[Raff(2021)]{Raff2020c}
Edward Raff.
\newblock {Research Reproducibility as a Survival Analysis}.
\newblock In \emph{The Thirty-Fifth AAAI Conference on Artificial
  Intelligence}, 2021.
\newblock URL \url{http://arxiv.org/abs/2012.09932}.

\bibitem[Redner(1998)]{Redner1998}
S~Redner.
\newblock {How popular is your paper? An empirical study of the citation
  distribution}.
\newblock \emph{The European Physical Journal B - Condensed Matter and Complex
  Systems}, 4\penalty0 (2):\penalty0 131--134, 1998.
\newblock ISSN 1434-6036.
\newblock \doi{10.1007/s100510050359}.
\newblock URL \url{https://doi.org/10.1007/s100510050359}.

\bibitem[Sculley et~al.(2015)Sculley, Holt, Golovin, Davydov, Phillips, Ebner,
  Chaudhary, Young, Crespo, and Dennison]{10.5555/2969442.2969519}
D~Sculley, Gary Holt, Daniel Golovin, Eugene Davydov, Todd Phillips, Dietmar
  Ebner, Vinay Chaudhary, Michael Young, Jean-Francois Crespo, and Dan
  Dennison.
\newblock {Hidden Technical Debt in Machine Learning Systems}.
\newblock In \emph{Proceedings of the 28th International Conference on Neural
  Information Processing Systems - Volume 2}, NIPS'15, pp.\  2503–2511,
  Cambridge, MA, USA, 2015. MIT Press.

\bibitem[Seabold \& Perktold(2010)Seabold and Perktold]{seabold2010statsmodels}
Skipper Seabold and Josef Perktold.
\newblock statsmodels: Econometric and statistical modeling with python.
\newblock In \emph{9th Python in Science Conference}, 2010.

\bibitem[Serra-Garcia \& Gneezy(2021)Serra-Garcia and Gneezy]{Serra-Garcia2021}
Marta Serra-Garcia and Uri Gneezy.
\newblock {Nonreplicable publications are cited more than replicable ones}.
\newblock \emph{Science Advances}, 7\penalty0 (21):\penalty0 eabd1705, 5 2021.
\newblock ISSN 2375-2548.
\newblock \doi{10.1126/sciadv.abd1705}.
\newblock URL
  \url{https://advances.sciencemag.org/lookup/doi/10.1126/sciadv.abd1705}.

\bibitem[Shockley(1957)]{Shockley1957}
William Shockley.
\newblock {On the Statistics of Individual Variations of Productivity in
  Research Laboratories}.
\newblock \emph{Proceedings of the IRE}, 45\penalty0 (3):\penalty0 279--290,
  1957.
\newblock ISSN 0096-8390.
\newblock \doi{10.1109/JRPROC.1957.278364}.
\newblock URL \url{http://ieeexplore.ieee.org/document/4056505/}.

\bibitem[Stegehuis et~al.(2015)Stegehuis, Litvak, and Waltman]{Stegehuis2015}
Clara Stegehuis, Nelly Litvak, and Ludo Waltman.
\newblock {Predicting the long-term citation impact of recent publications}.
\newblock \emph{Journal of Informetrics}, 9\penalty0 (3):\penalty0 642--657,
  2015.
\newblock ISSN 18755879.
\newblock \doi{10.1016/j.joi.2015.06.005}.

\bibitem[Sun et~al.(2020)Sun, Yu, Fang, Yang, Qu, Zhang, and
  Geng]{10.1145/3383313.3412489}
Zhu Sun, Di~Yu, Hui Fang, Jie Yang, Xinghua Qu, Jie Zhang, and Cong Geng.
\newblock {Are We Evaluating Rigorously? Benchmarking Recommendation for
  Reproducible Evaluation and Fair Comparison}.
\newblock In \emph{Fourteenth ACM Conference on Recommender Systems}, RecSys
  '20, pp.\  23–32, New York, NY, USA, 2020. Association for Computing
  Machinery.
\newblock ISBN 9781450375832.
\newblock \doi{10.1145/3383313.3412489}.
\newblock URL \url{https://doi.org/10.1145/3383313.3412489}.

\bibitem[Thelwall \& Wilson(2014)Thelwall and Wilson]{Thelwall2014a}
Mike Thelwall and Paul Wilson.
\newblock {Regression for citation data: An evaluation of different methods}.
\newblock \emph{Journal of Informetrics}, 8\penalty0 (4):\penalty0 963--971, 10
  2014.
\newblock ISSN 17511577.
\newblock \doi{10.1016/j.joi.2014.09.011}.
\newblock URL
  \url{https://linkinghub.elsevier.com/retrieve/pii/S1751157714000923}.

\bibitem[Traag(2021)]{Traag2021}
V~A Traag.
\newblock {Inferring the causal effect of journals on citations}.
\newblock \emph{Quantitative Science Studies}, 2\penalty0 (2):\penalty0
  496--504, 7 2021.
\newblock ISSN 2641-3337.
\newblock \doi{10.1162/qss{\_}a{\_}00128}.
\newblock URL \url{https://doi.org/10.1162/qss_a_00128}.

\bibitem[Varga(2019)]{Varga2019}
Attila Varga.
\newblock {Shorter distances between papers over time are due to more
  cross-field references and increased citation rate to higher-impact papers}.
\newblock \emph{Proceedings of the National Academy of Sciences of the United
  States of America}, 116\penalty0 (44):\penalty0 22094--22099, 2019.
\newblock ISSN 10916490.
\newblock \doi{10.1073/pnas.1905819116}.

\bibitem[Vul et~al.(2008)Vul, Harris, Winkielman, and Pashler]{Vul2008}
Edward Vul, Christine Harris, Piotr Winkielman, and Harold Pashler.
\newblock {Voodoo Correlations in Social Neuroscience}.
\newblock \emph{Perspectives on Psychological Science}, 2008.

\bibitem[Wallace et~al.(2009)Wallace, Larivi{\`{e}}re, and
  Gingras]{Wallace2009}
Matthew~L. Wallace, Vincent Larivi{\`{e}}re, and Yves Gingras.
\newblock {Modeling a century of citation distributions}.
\newblock \emph{Journal of Informetrics}, 3\penalty0 (4):\penalty0 296--303,
  2009.
\newblock ISSN 17511577.
\newblock \doi{10.1016/j.joi.2009.03.010}.

\bibitem[Yan et~al.(2011)Yan, Tang, Liu, Shan, and Li]{10.1145/2063576.2063757}
Rui Yan, Jie Tang, Xiaobing Liu, Dongdong Shan, and Xiaoming Li.
\newblock {Citation Count Prediction: Learning to Estimate Future Citations for
  Literature}.
\newblock In \emph{Proceedings of the 20th ACM International Conference on
  Information and Knowledge Management}, CIKM '11, pp.\  1247–1252, New York,
  NY, USA, 2011. Association for Computing Machinery.
\newblock ISBN 9781450307178.
\newblock \doi{10.1145/2063576.2063757}.
\newblock URL \url{https://doi.org/10.1145/2063576.2063757}.

\end{thebibliography}
\bibliographystyle{iclr2022_conference}

\appendix
\section{Appendix}

We note that we have released a modified version of the ML citation data at the URL specified. In initial release of the data we kept paper titles withheld due to concern that marking nearly 100 papers as “not reproducible by this author in their attempts” would be misconstrued as “not reproducible”, cause potentially high stress for junior authors in the list, and meet with potential acrimony at large. While we believe time may have cooled some concerns, the large number of initial emails, sometimes heated, about the work makes us fear that such self-censorship was unfortunately the best choice of action. As such we are striking a balance that in releasing a version of the data with citations by year, would be too trivially easy to determine the entire author list. 

As such a small amount of noise has been added such that the results are generally identical in re-running the analysis, but keeps the reverse of the names at least not completely trivial. Our hope was to use differential privacy to perform a more robust release, but the nature of citation count data meant that the amount of required noise had a hugely detrimental impact on the results that prevented any replication. 

We hope the reader will understand that balance we are trying to make, and that the data is still useful.

\end{document}